\documentclass{elsarticle}

\usepackage{float}
\usepackage{graphicx}
\usepackage{subcaption}
\usepackage{algorithm2e}
\usepackage{siunitx}
\usepackage{amsmath}
\usepackage{physics,bm}
\usepackage{nomencl}
\usepackage{numcompress}

\usepackage{hyperref}

\graphicspath{ {./images/} }

\begin{document}

\begin{frontmatter}

\title{A diffuse interface method for simulation-based screening of heat transfer processes with complex geometries}

\author[chemeng]{Elizabeth J. Monte}

\author[chemeng]{James Lowman}

\author[chemeng,physics]{Nasser Mohieddin Abukhdeir\corref{mycorrespondingauthor1}}
\cortext[mycorrespondingauthor1]{Corresponding author, nmabukhdeir@uwaterloo.ca}
\ead[url]{https://uwaterloo.ca/comphys}

\address[chemeng]{Department of Chemical Engineering}
\address[physics]{Department of Physics and Astronomy\\University of Waterloo\\200 University Avenue West\\Waterloo, Ontario, Canada N2L 3G1}

\begin{abstract}
    Frequently, the design of physicochemical processes requires screening of large numbers of alternative designs with complex geometries.
    These geometries may result in conformal meshes which introduce stability issues, significant computational complexity, and require user-interaction for their creation.
    In this work, a method for simulation of heat transfer using the diffuse interface method to capture complex geometry is presented as an alternative to a conformal meshing, with analysis and comparisons given.
    The methods presented include automated non-iterative generation of phase fields from CAD geometries and an extension of the diffuse interface method for mixed boundary conditions.
    Simple measures of diffuse interface quality are presented and found provide predictions of performance. The method is applied to a realistic heat transfer problem and compared to the traditional conformal mesh approach.
    It is found to enable reasonable accuracy at an order-of-magnitude reduction in simulation time or comparable accuracy for equivalent simulation times.
\end{abstract}


\end{frontmatter}


\section{Introduction}

Computational multiphysics simulations are a key enabler for the advancement of the design and optimization of physicochemical processes.
Design, control, and optimization of these processes requires detailed insight, and low-cost nonhazardous methods for the evaluation of candidate designs, for which simulation-based screening is ideal.
However, several challenges posed by multiphysics simulations of industrially relevant processes, scales, and geometries exist such as computational complexity and numerical stability (\cite{Slotnick2014}).

Industrial processes involve geometries with both complex topology and significantly varying length scales.
In order to perform computational multiphysics simulations of these processes, standard spatial discretization involves the use of unstructured \emph{conformal meshes}, which capture domain shape through conforming to its boundary.
This typically results in spatial discretization length scales being far smaller than those characteristic to the underlying physicochemical phenomena, but which are required to capture boundary curvature.
Consequently, the numerical solutions of multiphysics models using conformal meshes can suffer from stability issues, \textit{eg.} Courant number constraints for convection-dominated processes, and have significant additional computational complexity compared to structured meshes of comparable size.
Also, the computational cost and frequent need for user interaction in the creation of conformal meshes (\cite{Franz2012}) are detrimental to the evaluation of large numbers of varied geometries as part of a design process.
This motivates the use of alternative methods for mesh generation and, more generally, the imposition of complex domain boundaries and boundary conditions (\cite{Anderson1998,Scardovelli1999,Mittal2005}).

There exist many approaches for approximating complex boundaries and interfaces as an alternative to conformal meshing.
In general, these methods involve the creation of a \emph{nonconformal mesh} and the imposition of boundary conditions through the addition of various types of forcing terms within a subdomain of the nonconformal mesh that conforms to the desired complex geometry.
These methods have been classified in the literature depending on the type of boundary or interface being modelled: solid/solid, fluid/solid or fluid/fluid.
\emph{Immersed boundary methods} are those which were developed for solid/solid and fluid/solid interfaces (\cite{Mittal2005}).
\emph{Diffuse-interface methods} are those which were developed for fluid/fluid interfaces (\cite{Anderson1998}).
There is significant overlap and interchange of this classification and so-called continuous-forcing immersed boundary methods (\cite{Mittal2005}) and diffuse-interface methods are similar in approach.

Over the past decade, substantial progress has been made with respect to the accuracy, stability, and application of these types of methods (\cite{Nguyen2017}).
\cite{Ramiere2007} developed a continuous immersed boundary method for mixed boundary conditions (Dirichlet, Neumann, Robin).
\cite{Li2009} developed different continuous immersed boundary condition approximations (Dirichlet, Neumann, and Robin) compatible with a moving diffuse interface, and showed that the resulting solutions converge to the conformal boundary condition in the sharp interface limit.
This was accomplished through the use of the method of matched asymptotic expansions.
\cite{Lervag2014} performed an asymptotic analysis of Neumann and Robin boundary conditions on a steady reaction-diffusion problem and compared the accuracy of different diffuse interface formulations.
Results showed that the boundary conditions proposed in \cite{Li2009} are either first order or second order accurate with respect to diffuse interface length scale, depending on which formulation is used.
\cite{Schlottbom2016} examined the diffuse interface on elliptic problems with Dirichlet boundary conditions, showing rates of convergence to the conformal solution for uniform and adaptive mesh refinement (local to the interface).
\cite{Franz2012} also showed the convergence of the Dirichlet boundary conditions formulated by \cite{Li2009}, similar to the results presented in \cite{Lervag2014}, and again applied these methods to the solution of a steady reaction-diffusion equation.

The stability of diffuse interface methods has been advanced in the past decade, particularly the stability of Dirichlet boundary conditions (\cite{Juntunen2009, Nguyen2018}), through the use of the Nitsche method (\cite{Nitsche1971}).
The Nitsche method provides a stable variationally consistent method (\cite{Nguyen2018,Evans2013}) for weakly enforcing Dirichlet boundary conditions by incorporating flux terms into the weak form of the differential equation to penalize deviations from those boundary conditions.
\cite{Nguyen2018} show that a single parameter introduced in the Nitsche method should be chosen as the largest eigenvalue for that system.
\cite{Schillinger2016} have recently developed a non-symmetric Nitsche method, which is parameter-free, which avoids potential stability issues resulting from using extreme values for the stabilization parameter.

The diffuse interface method has been applied to a broad range of applications.
\cite{Teigen2009} applied the method to modelling diffusion and adsorption/desorption within a deformable interface using both adaptive mesh refinement and a multigrid method.
\cite{Aland2010} used the diffuse interface method with the Cahn-Hilliard/Navier-Stokes model for immiscible two-phase flow in complex geometries, accounting for contact lines.
\cite{Nguyen2017} used the diffuse interface method to perform stress analysis of bone structures directly from tomography data of bone mineral density.
They formulated the diffuse interface through artificial diffusion of three-dimensional tomography data and compared their results to standard voxel finite-cell methods, finding similar accuracy.
\cite{Stoter2017} applied this approach to the simulation of single-phase/porous media transport within the human liver using MRI imaging.
However, they found that for equivalent accuracy, using a traditional conformal mesh approach was less computationally complex.
Recently, \cite{Treeratanaphitak2021} applied the diffuse interface method to the solution of the two-fluid model for liquid/dispersed gas flows, finding that the kernels used for constructing the diffuse-interface had negligible effect on accuracy for interface scales significantly smaller than the geometry scale.

While significant progress has been made, there are still significant practical challenges to the application of diffuse interface methods for engineering design.
Firstly, computation of a phase field corresponding to a geometry specified through CAD software is a non-trivial task.
Methods have been developed for this task (\cite{Nguyen2017, Nguyen2018, Teigen2009, Aland2010, Stoter2017, Aland2020}), but all involve the solution of a diffusion equation to generate a smooth differentiable phase field or other computationally complex algorithms.
Additionally, physicochemical processes rarely involve a single uniform boundary condition and, instead, typically involve varying boundary conditions over sub-domains of the conformal surface.
Yet, most applications of diffuse interface methods have been limited to the uniform boundary condition case.
\cite{Li2009} presented mathematical formulations for Dirichlet, Neumann, and Robin boundary conditions, but only applied a single boundary condition in each of their test cases.
\cite{Lervag2014} tested spatially varying Neumann and Robin boundary conditions, but did not combine different types of boundary conditions in the same simulation.
\cite{Ramiere2007} combined Dirichlet, Neumann, and Robin boundary conditions in their test cases.
However, the Neumann boundary conditions were applied to portions of the mesh which conform to the complex geometry boundary, not to areas approximated with a diffuse interface.
They did, in later work, combine Dirichlet and Robin boundary conditions over different sections of a diffuse interface boundary approximation, although it is unclear how the specific boundary sections were isolated (\cite{Ramiere2007a}).

Within the aforementioned context, the overall objective of this work is to develop a comprehensive diffuse interface method for use in screening heat transfer processes with complex geometries.
That is, the intent is not to replace conformal mesh simulations but instead to reduce the computational complexity during the screening process through approximation of the geometries and boundary conditions.
Specific objectives include:
\begin{enumerate}
    \item{Development of an automated and non-iterative method for the generation of phase fields from CAD descriptions of geometries, including segmenting different boundary conditions.}
    \item{Development of a \textit{stable} method for imposing mixed boundary conditions using the diffuse interface method.}
    \item{Validation of the developed methods with the solution of a heat transfer problem with spatially-varying boundary conditions on a complex geometry and evaluate performance compared to the reference/conformal mesh solution.}
\end{enumerate}

The paper is organized as follows: First, the diffuse interface and Nitsche methods are reviewed and demonstrated for Poisson's equation in Section \ref{sec:background}.
Then, the method for generation of phase fields from CAD geometries and quality measures for predicting its performance are presented and demonstrated in Sections \ref{sec:results:generation}-\ref{sec:results:metrics}.
A method for imposing mixed boundary conditions for the diffuse interface is then presented in Section \ref{sec:results:boundary} and applied to a heat transfer problem (LED heat sink) in Section \ref{sec:results:validation}.
Finally, conclusions are summarized in Section \ref{sec:conclusions}.

\section{Model Formulation}
\label{sec:background}

A large and diverse amount of past research has been devoted to phase field methods.
Research on and applications of phase field methods has mainly focused either on fluid-structure interactions (\cite{Mittal2005}), or fluid-fluid interactions near and far from critical points (\cite{Anderson1998}).
Some previous work has also examined phase field methods for use on a single domain, particularly \cite{Nguyen2018}.
In this work, the use of a stationary diffuse interface is focused on and this section is limited to relevant background.

\subsection{Diffuse Interface Method}
\label{sec:background:DIM}

\begin{figure}[H]
	\centering
	\begin{subfigure}[b]{0.24\textwidth}
		\includegraphics[width=\linewidth]{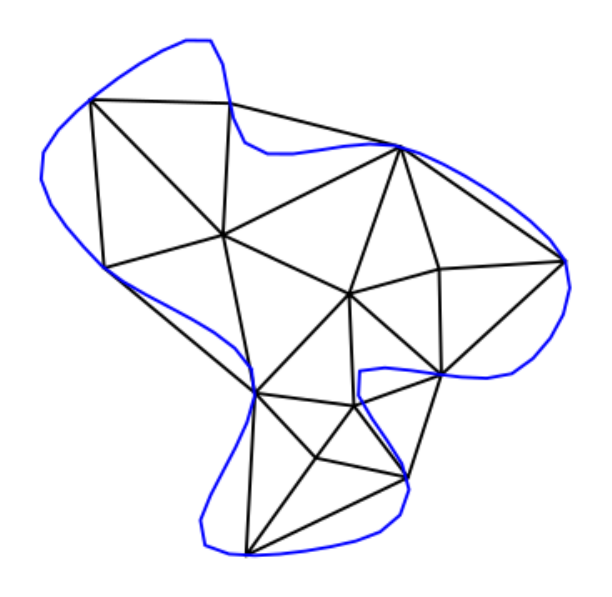}
		\caption{}
	\end{subfigure}%
	\hfill
	\begin{subfigure}[b]{0.24\textwidth}
		\includegraphics[width=\linewidth]{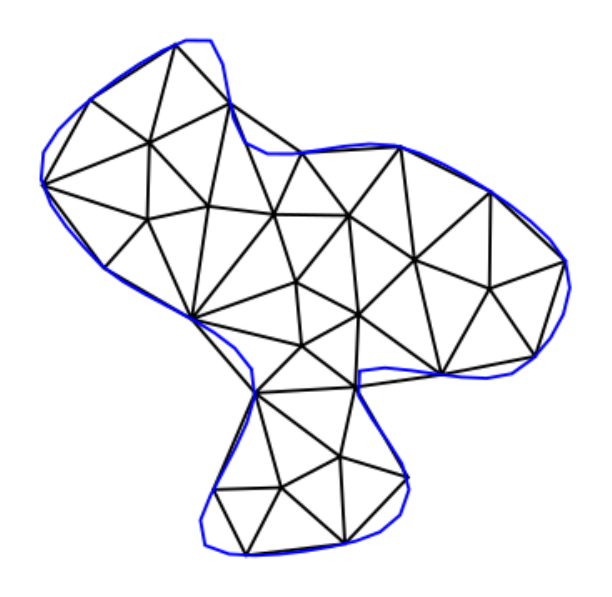}
		\caption{}
	\end{subfigure}%
	\hfill
	\begin{subfigure}[b]{0.24\textwidth}
		\includegraphics[width=\linewidth]{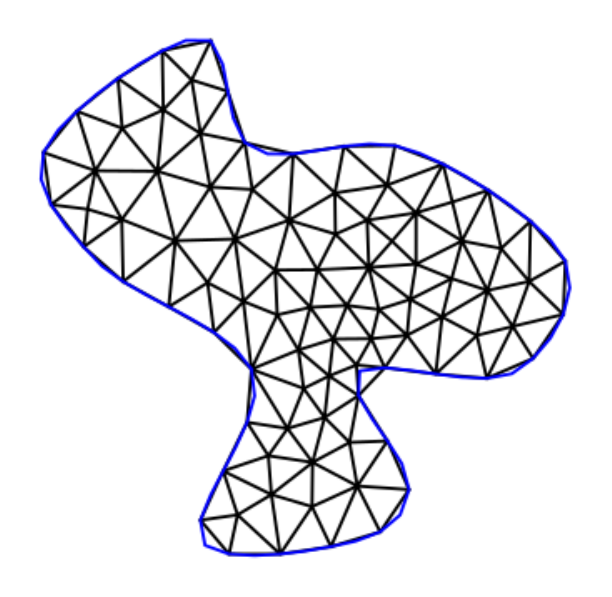}
		\caption{}
	\end{subfigure}%
	\hfill
	\begin{subfigure}[b]{0.24\textwidth}
		\includegraphics[width=\linewidth]{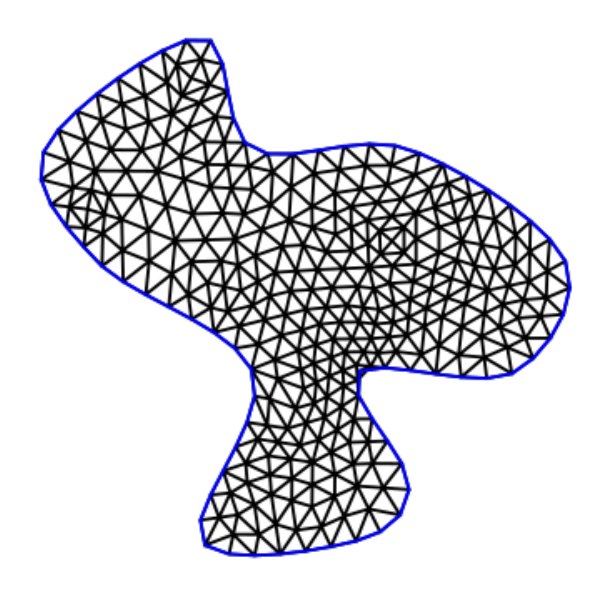}
		\caption{}
	\end{subfigure}%
	\caption{An example complex geometry meshed by conformal meshes with increasing levels of refinement. A high level of refinement is needed for mesh edges to reasonably conform to the geometry boundary, particularly in regions of high curvature. \label{fig:intro:overview_2}}
\end{figure}

For a given problem, some complex geometry $\Omega$ is defined as the simulation domain.
The standard conformal meshing approach would be to discretize $\Omega$ with an unstructured mesh which conforms to its boundary ($\partial \Omega$), shown in Figure \ref{fig:intro:overview_2}.
As can be seen, this approach to capturing domain geometry does not, in general, exactly conform apart from boundary nodes of the mesh elements.
Conformance increases as spatial discretization scale decreases and converges to the exact geometry as this scale approaches zero.
Additionally, generation of conformal meshes typically requires them to be unstructured, involves significant computational complexity, and significantly increases the complexity of mesh partitioning for parallel computation.

\begin{figure}[H]
	\centering
	\begin{subfigure}[b]{0.24\textwidth}
		\includegraphics[width=\linewidth]{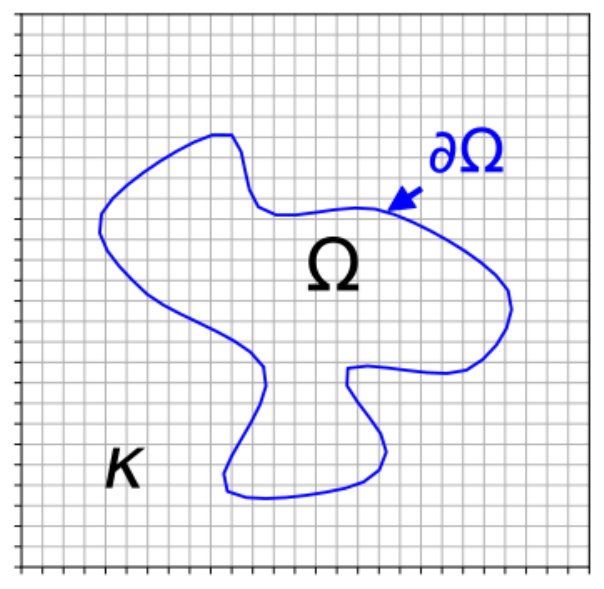}
		\caption{}
	\end{subfigure}%
	\hfill
	\begin{subfigure}[b]{0.24\textwidth}
		\includegraphics[width=\linewidth]{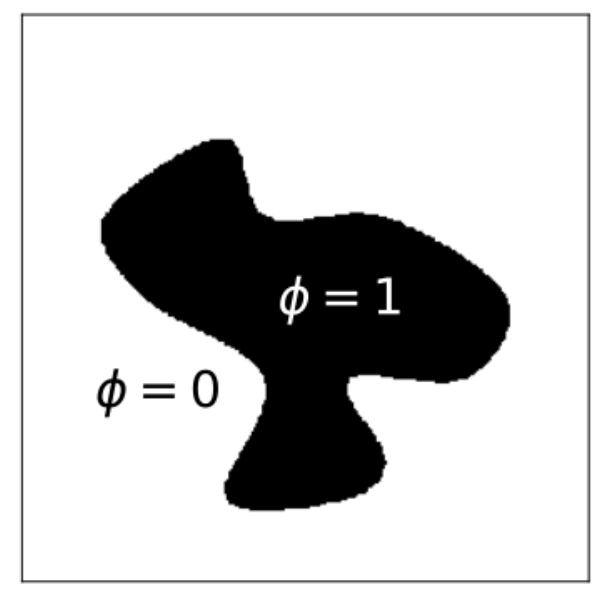}
		\caption{}
	\end{subfigure}%
	\hfill
	\begin{subfigure}[b]{0.24\textwidth}
		\includegraphics[width=\linewidth]{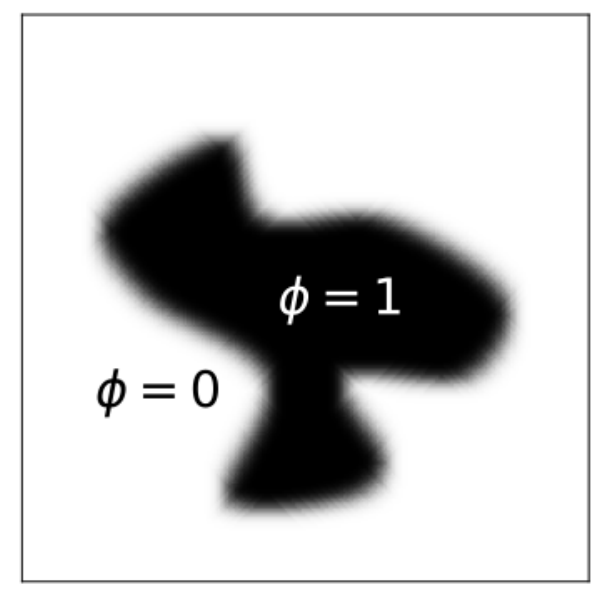}
		\caption{}
	\end{subfigure}%
	\hfill
	\begin{subfigure}[b]{0.24\textwidth}
		\includegraphics[width=\linewidth]{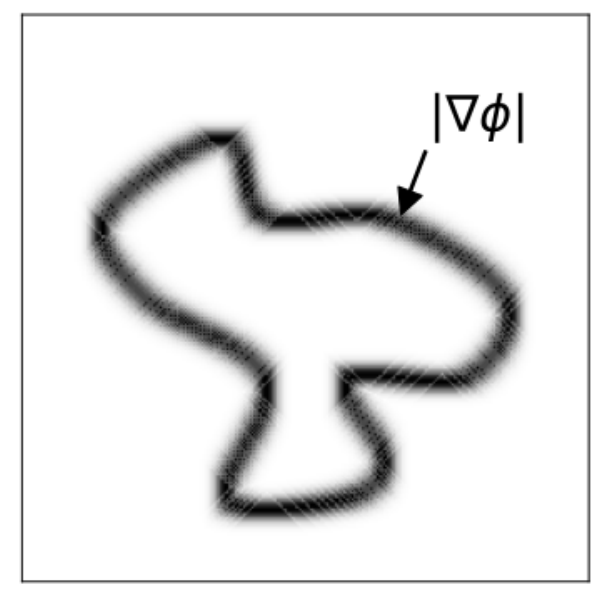}
		\caption{}
	\end{subfigure}\\
    \begin{subfigure}[b]{0.49\textwidth}
		\includegraphics[width=\linewidth]{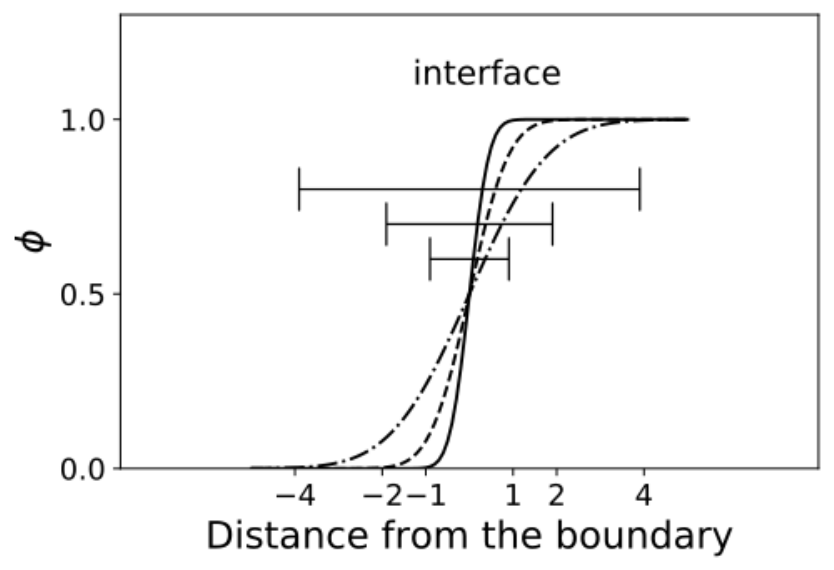}
		\caption{}
	\end{subfigure}%
	\hfill
	\begin{subfigure}[b]{0.49\textwidth}
		\includegraphics[width=\linewidth]{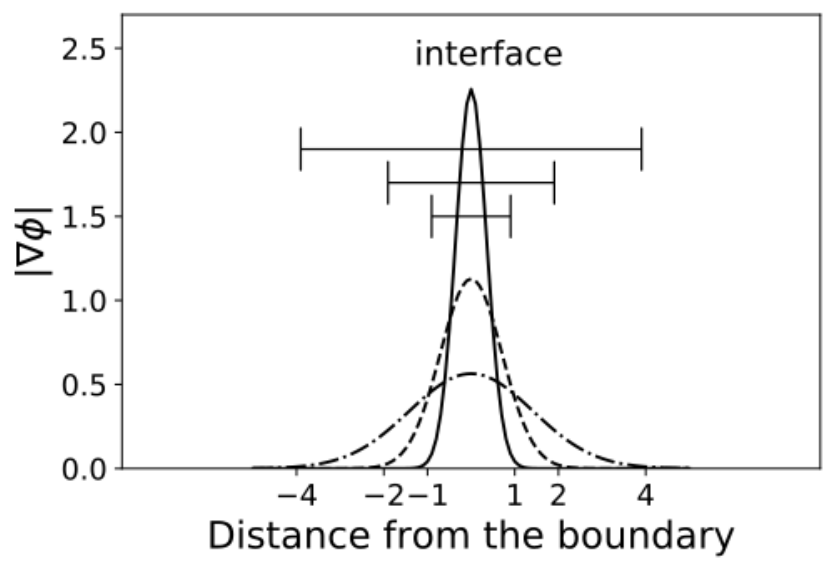}
		\caption{}
	\end{subfigure}%
	\caption{An example complex geometry approximated by the diffuse interface method with (a) the geometry encompassed by a structured mesh (mesh coarsened for visibility), (b) the phase field approximating the geometry on a refined structured mesh, (c) the same phase field now with a diffuse interface, (d) the diffuse approximation of the interface indicated by $\lvert \nabla \phi \rvert$, and further schematics of (e) the phase field and (f) its gradient for different diffuse interface widths. Based on Figures 1, 3, and 4 in \cite{Nguyen2018}.
\label{fig:intro:overview}}
\end{figure}

Alternatively, the diffuse interface method defines a domain $\kappa$, typically Cartesian (rectangular cuboid), which encompasses $\Omega$ and is then discretized without conforming to the boundaries $\partial \Omega$, as shown in Figure \ref{fig:intro:overview}a.
A structured mesh may be used, which requires significantly less computational complexity compared to unstructured meshes required by the conformal approach.
In order to capture the desired original geometry, a scalar phase field $\phi$ is defined on $\kappa$.
The phase field has the value of one within $\Omega$, the value zero outside of $\Omega$, and may be discontinuous (sharp-interface) or continuous (diffuse interface) at the interface between the two subdomains, shown in Figures \ref{fig:intro:overview}b-c, respectively.
The sharp boundary of the original geometry can be approximated by the level set $\phi = 0.5$, while the diffuse approximation extends over the region $\lvert \nabla \phi \rvert > 0$ (Figure \ref{fig:intro:overview}d). This region approaches the Dirac delta function $\delta$ as the diffuse interface width approaches zero, the sharp-interface limit, as shown in Figures \ref{fig:intro:overview}e-f.

Given a diffuse interface approximation of a domain $\Omega$, the weak form of the well-posed problem may be reformulated from a conformal to diffuse interface (nonconformal) form using the following integral identities (\cite{Nguyen2018}),
\begin{align}
\label{eqn:diffuse1}
\int_{\Omega} A \: d\Omega &= \int_{\kappa} H A \: d\kappa \approx \int_{\kappa} \phi A \: d\kappa \\
\label{eqn:diffuse2}
\int_{\partial \Omega} B \: d\Omega &= \int_{\kappa} \delta_{\partial \Omega} B \: d\kappa \approx \int_{\kappa} \lvert \bm{\nabla} \phi \rvert B \: d\kappa
\end{align}
where the phase field and its gradient approximate the Heaviside function $H$ and the Dirac delta function $\delta_{\partial \Omega}$ respectively.
Functions $A$ and $B$ can represent scalar or vector fields.
The unit normal vector to the complex domain boundary can also be approximated as follows (\cite{Nguyen2018}),
\begin{equation}
\label{eqn:diffuse3}
\bm{n} \approx \frac{-\bm{\nabla} \phi}{\lvert \bm{\nabla} \phi \rvert}
\end{equation}

This approach is demonstrated using the Poisson problem with generalized boundary conditions and formulated for the finite element method,
\begin{align}
\label{eqn:poisson1}
-\nabla^2 u &= f \mbox{ on } \Omega \\
\label{eqn:poisson2}
u &= h \mbox{ on } \partial \Omega_D \\
\label{eqn:poisson3}
-\bm{n} \cdot \bm{\nabla} u &= g \mbox{ on } \partial \Omega_N \\
\label{eqn:poisson4}
-\bm{n} \cdot \bm{\nabla} u &= r(u - q) \mbox{ on } \partial \Omega_R
\end{align}
where $\partial \Omega_D$, $\partial \Omega_N$, and $\partial \Omega_R$ denote the boundary sections on which Dirichlet, Neumann, and Robin boundary conditions are applied respectively.
By introducing a test (or weighting) function $v$ in the Sobolev space $H^1 (\Omega)$, the weak form can be formulated,
\begin{align}
\label{eqn:weak1}
\int_{\Omega} (\nabla^2 u + f)v \: d\Omega &= 0 \\
\label{eqn:weak2}
\int_{\Omega} \bm{\nabla} u \cdot \bm{\nabla} v \: d\Omega - \int_{\Omega} \bm{\nabla} \cdot (v\bm{\nabla}u) \: d\Omega &= \int_{\Omega} fv \: d\Omega \\
\label{eqn:weak3}
\int_{\Omega} \bm{\nabla} u \cdot \bm{\nabla} v \: d\Omega - \int_{\partial \Omega} v (\bm{n} \cdot \bm{\nabla} u) \: d\partial \Omega &= \int_{\Omega} fv \: d\Omega
\end{align}
where integration by parts and the divergence theorem are used.

Neumann and Robin boundary conditions are imposed through substitution into the surface integral term,
\begin{align}
\label{eqn:weak5}
\int_{\Omega} \bm{\nabla} u \cdot \bm{\nabla} v \: d\Omega + \int_{\partial \Omega_N} v g \: d\partial \Omega &= \int_{\Omega} fv \: d\Omega \\
\label{eqn:weak6}
\int_{\Omega} \bm{\nabla} u \cdot \bm{\nabla} v \: d\Omega + \int_{\partial \Omega_R} v r(u-q) \: d\partial \Omega &= \int_{\Omega} fv \: d\Omega
\end{align}
Dirichlet boundary conditions are imposed by imposing both $u = h$ and $v = 0$ on $\partial \Omega_D$.

The diffuse interface method transforms the volume integrals on $\Omega$ and the surface integrals on $\partial \Omega$ into volume integrals on $\kappa$.
Eqns. \ref{eqn:diffuse1}-\ref{eqn:diffuse3} can be inserted into eqns. \ref{eqn:weak5}-\ref{eqn:weak6} to produce the diffuse interface formulation for Neumann and Robin boundary conditions,
\begin{align}
\label{eqn:diffuse5}
&\int_{\kappa} \bm{\nabla} u \cdot \bm{\nabla} v \phi \: d\kappa + \int_{\kappa} v g \lvert \bm{\nabla} \phi \rvert \: d\kappa = \int_{\kappa} fv \phi \: d\kappa \\
\label{eqn:diffuse6}
&\int_{\kappa} \bm{\nabla} u \cdot \bm{\nabla} v \phi \: d\kappa + \int_{\kappa} v r(u-q) \lvert \bm{\nabla} \phi \rvert \: d\kappa = \int_{\kappa} fv \phi \: d\kappa
\end{align}
Dirichlet boundary conditions require additional treatment, as will be discussed in the following section.

Since $\phi$ goes to zero outside of the approximation to the complex geometry, singularities are likely if $\kappa$ is much larger than $\Omega$.
This can be prevented by redefining the phase field as follows,
\begin{align}
\phi' = \alpha + (1 - \alpha) \phi
\end{align}
where $\phi'$ is the new phase field and $\alpha$ is some small constant \textit{eg.} $\num{1e-8}$.
Alternatively, any mesh elements in $\kappa$ on which $\phi$ is zero can be removed before solving (\cite{Nguyen2018}).

\subsection{Stabilization of Dirichlet Conditions -- The Method of Nitsche}
\label{sec:background:Nitsche}

The diffuse interface method requires additional treatment of Dirichlet boundary conditions due to the lack of an explicit conformal boundary.
Most previous work on phase field methods has focused on penalty methods or variants of the Nitsche method (\cite{Nitsche1971}).

The penalty method directly penalizes deviations from the Dirichlet boundary condition at the diffuse boundary by adding the term (\cite{Juntunen2009}),
\begin{equation}\label{eqn:penalty1}
    \beta \int_{\partial \Omega_D} v (u - h) \: d \partial \Omega_D
\end{equation}
to the mathematical formulation, where $\beta$ is a stabilization parameter which is problem dependent.
\cite{Li2009b} show several variants of the penalty method; \cite{Schlottbom2016} and \cite{Ramiere2007} also use penalty-type methods.
However, the Nitsche method (\cite{Nitsche1971}) has been recently shown to enable higher accuracy and less sensitivity to the choice of a stabilization parameter compared to the penalty method (\cite{Nguyen2018}).
The Nitsche method both penalizes deviation from the Dirichlet boundary condition and includes additional terms to maintain symmetry (\cite{Benk2012}),
\begin{equation}
\begin{split}\label{eqn:nitsche1}
    \int_{\Omega} \bm{\nabla} u \cdot \bm{\nabla} v \: d\Omega - \int_{\partial \Omega_D} u (\bm{n} \cdot \bm{\nabla} v) \: d\partial \Omega_D - \int_{\partial \Omega_D} v (\bm{n} \cdot \bm{\nabla} u) \: d\partial \Omega_D \\
    + \: \beta \int_{\partial \Omega_D} v (u - h) \: d\partial \Omega_D = \int_{\Omega} fv \: d\Omega - \int_{\partial \Omega_D} h (\bm{n} \cdot \bm{\nabla} v) \: d\partial \Omega_D
\end{split}
\end{equation}
where $\beta$ is a stabilization parameter which is problem dependent.

The Nitsche method was first applied within the phase field method by \cite{Freund1995} and has since been applied in other studies (\cite{Nguyen2017,Nguyen2018,Stoter2017,Benk2012}).
Using eqns. \ref{eqn:diffuse1}-\ref{eqn:diffuse3}, the diffuse interface formulation incorporating the Nitsche method becomes,
\begin{equation}
\begin{split}\label{eqn:nitsche2}
    \int_{\kappa} \bm{\nabla} u \cdot \bm{\nabla} v \phi \: d\kappa + \int_{\kappa} u (\bm{\nabla} \phi \cdot \bm{\nabla} v) \: d\kappa + \int_{\kappa} v (\bm{\nabla} \phi \cdot \bm{\nabla} u) \: d\kappa \\
    + \: \beta \int_{\kappa} v (u - h) \lvert \bm{\nabla} \phi \rvert \: d\kappa = \int_{\kappa} fv \phi \: d\kappa + \int_{\kappa} h (\bm{\nabla} \phi \cdot \bm{\nabla} v) \: d\kappa \\
\end{split}
\end{equation}
The Nitsche method has also been extended to Neumann and Robin boundary conditions by \cite{Juntunen2009}.

As shown in eqns. \ref{eqn:nitsche1}-\ref{eqn:nitsche2}, the Nitsche method introduces a new parameter $ \beta $, generally defined as $ \beta = \gamma/h$ (\cite{Juntunen2009, Evans2013, Benk2012, Hansbo2005, Ruess2013, Ruess2014, Prenter2018}).
The characteristic length of each mesh element $ h $, \textit{eg.} the average or maximum scale of the mesh element.
With a structured mesh, such as those used by the diffuse interface method, $h$ is constant over the entire mesh, but an unstructured mesh would require $h$ to be defined element-wise (\cite{Prenter2018}).
A stability constant $ \gamma $, must be large enough to satisfy a stability inequality, see \cite{Hansbo2005} for further discussion.
\cite{Evans2013} show that $ \gamma $ should be proportional to the square of the numerical solution interpolant order.
\cite{Benk2012}, \cite{Ruess2013}, and \cite{Ruess2014}, further find that $\gamma$ values of $2-10$, $10-100$, and $100-1000$ respectively give reasonable accuracy for interpolant orders of 1-3.
This work then uses the following equation for $\beta$,
\begin{equation}\label{eqn:nitsche4}
    \beta = \frac{10n^2}{\Delta x}
\end{equation}
where $n$ refers to the interpolant order of the numerical solution and the mesh spacing $\Delta x$ is used for $h$.
$\beta$ could also be calculated element-wise as the maximum eigenvalue of the stability inequality, see, for example, (\cite{Nguyen2018, Prenter2018}).

\section{Results and Discussion}

The two complementary methods developed in this work, within the context of enabling the use of the diffuse interface method for screening of heat transfer processes, are:
\begin{enumerate}
  \item An automated and non-iterative method for generating phase fields from CAD geometries, including segmenting different boundary conditions.
  \item A method for imposing mixed boundary conditions using phase fields.
\end{enumerate}
These methods will be presented and validated using simple variations of steady-state diffusion equations with uniform and mixed Dirichlet and Neumann boundary conditions.

The first validation case is defined on a circular domain ($\tilde R = 1$) and was chosen due to the smoothness of the geometry boundary,
\begin{equation}\label{eqn:res:circle}
    -\nabla^2 T = \left(\frac{2 \pi}{\alpha} \right)^2 r^2 \sin\left({\frac{\pi r^2}{\alpha}}\right) - \frac{4 \pi}{\alpha} \cos\left({\frac{\pi r^2}{\alpha}}\right)
\end{equation}
with an exact solution,
\begin{equation}\label{eqn:res:circle_exact}
T = \sin\left({\frac{\pi r^2}{\alpha}}\right)
\end{equation}
for compatible Dirichlet and Neumann boundary conditions.
The parameter $\alpha$ is set to 0.05.

The second validation case is defined on a rectangular domain ($\tilde L = \tilde W = 1$),
\begin{equation}\label{eqn:res:square}
    -\nabla^2 T = 8 \pi^2 \sin\left({2\pi x}\right)\cos\left({2\pi y}\right)
\end{equation}
with an exact solution,
\begin{equation}\label{eqn:res:square_exact}
    T =  \sin\left({2\pi x}\right)\cos\left({2\pi y}\right)
\end{equation}
for compatible Dirichlet and Neumann boundary conditions.
This rectangular domain has sharp corners, which introduces more challenging geometric conditions for the imposition of a (continuous) diffuse interface approximation to the boundary.

\begin{figure}[H]
	\centering
	\begin{subfigure}[b]{0.35\textwidth}
		\includegraphics[width=\linewidth]{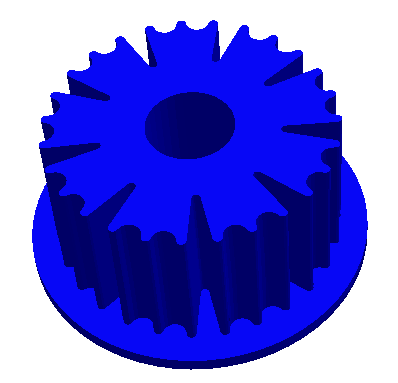}
		\caption{}
	\end{subfigure}%
	\begin{subfigure}[b]{0.55\textwidth}
		\includegraphics[width=\linewidth]{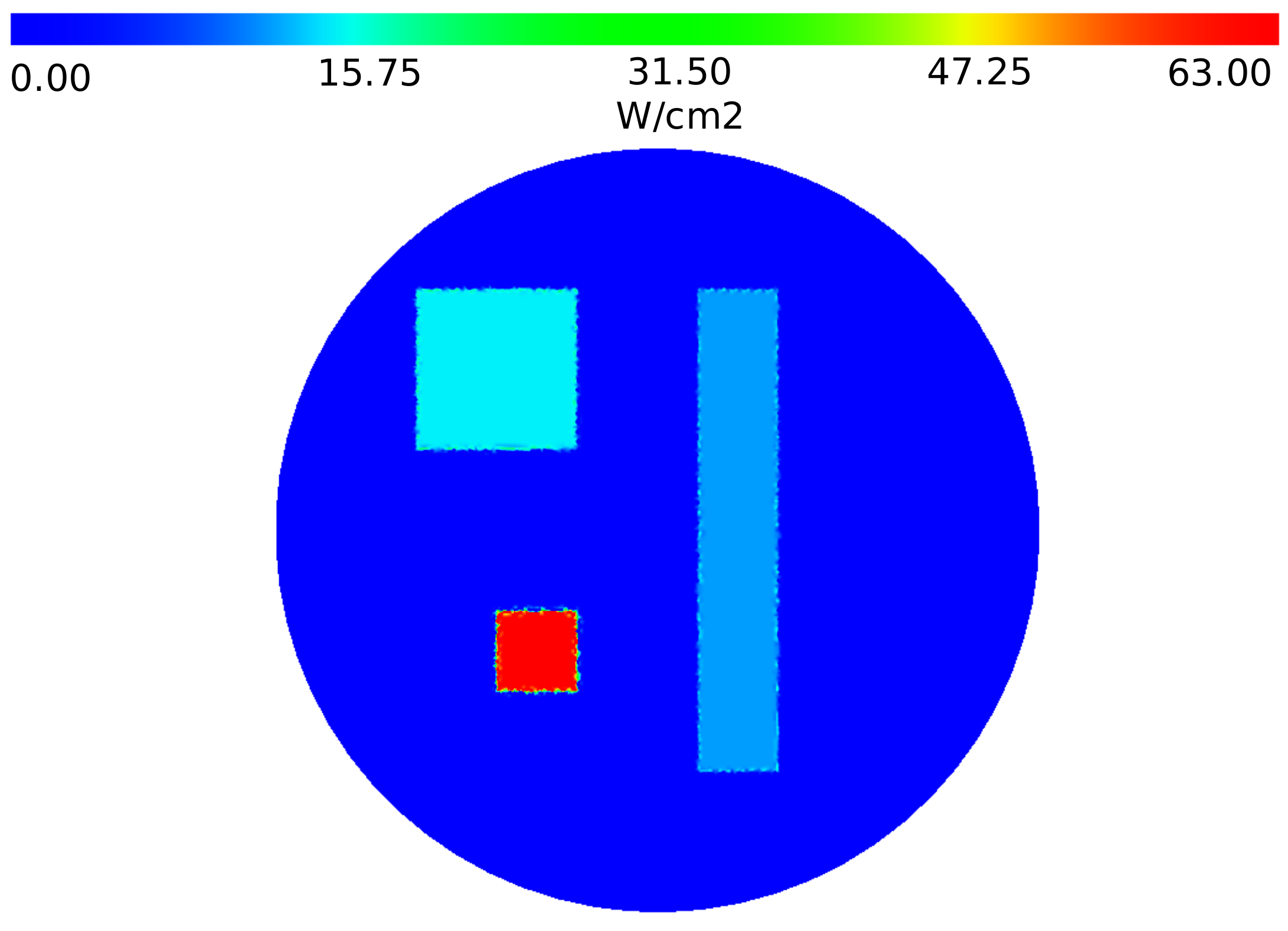}
		\caption{}
	\end{subfigure}%
	\caption{Example of a heat sink with a complex shape with (a) top-view in contact with the environment and (b) bottom-view in contact with a spatially varying heat source (in this case, several LEDs). CAD is based on \cite{Ghosh2015} with modifications.\label{fig:results:heat_sink}}
\end{figure}

Finally, the methods developed in this work were applied to a realistic heat transfer application, evaluation of the LED heat sink design shown in Figure \ref{fig:results:heat_sink}.
This is a diffusion problem with mixed Neumann (see Figure \ref{fig:results:heat_sink}b) and Robin boundary conditions, solved on a complex three dimensional geometry. It demonstrates the performance benefits of the diffuse interface method compared to a conformal mesh solution.

All simulations detailed in the following sections used the Netgen/NGSolve (\cite{ngsolve}) implementation of the finite element method.
Conformal meshing of the complex geometries used Netgen/NGSolve algorithms for automated mesh generation and defect removal.
Algorithms to generate and mark the phase field were implemented in Python and have been provided as supplemental material.

\subsection{Non-iterative method for continuous diffuse phase field generation} \label{sec:results:generation}

Past research applying phase field methods to scientific and engineering applications (\cite{Nguyen2017,Ramiere2007,Ramiere2007a,Nguyen2018,Teigen2009,Aland2010,Stoter2017,Aland2020,Benk2012,De2016,Mo2018}) has leveraged the ease of capturing complex, potentially deforming, geometries where a conformal mesh would be computationally infeasible.
In general, these past methods involve the projection of a binary discontinuous phase field, created from the geometry, onto the computational grid or mesh.
The binary phase value is then used to distinguish the inside/outside of the geometry as a sharp-interface.
However, this results in a discontinuity at the interface, shown in Figure \ref{fig:intro:overview}b, which is suitable for the application of discontinuous immersed boundary methods (\cite{Mittal2005}), but not for continuous immersed boundary methods.

In general, a continuous phase field can be generated from the discontinuous phase field (see Figures \ref{fig:intro:overview}c-d) through the solution of a diffusion equation or the use of a signed distance function.
\textit{Ad hoc} ``blurring'' has been used in immersed boundary methods to spread and interpolate the forcing functions (for example, \cite{TojaSilva2014}) and could also be applied to the diffuse interface method.
However, the continuous phase fields produced by blurring methods do not have easily quantifiable diffuse interface thicknesses, nor a uniform thickness throughout the boundary, so it becomes challenging to enforce the quality of the interface approximation.
Diffusion equation approaches solve the Allen-Cahn equation over the domain, using the discontinuous phase field as the initial condition and specifying an interface length scale parameter to control the width of the resulting phase field (for example, \cite{Nguyen2018}).
This phase field varies continuously from zero to one, behaving as the hyperbolic tangent function within the interface region.
However, these methods require iterating through a series of pseudo-timesteps to reach the metastable diffused state and additional computational complexity is required for numerical differentiation (assuming an explicit integration method is used).
Since the metastable solution of the Allen-Cahn equation is known as a function of the distance from the initial discontinuous interface, this iterative procedure could be avoided if such a signed distance function were known.
Several works have suggested the use of a signed distance function (\cite{Franz2012,Li2009,Lervag2014,Schlottbom2016,Teigen2009}), but have not suggested a method for generating this signed distance function when it is not known analytically.
\cite{Aland2020} provide a method for computing the signed distance function of any geometry from its surface triangulation.
However, their use of unstructured meshes necessitates the use of costly ray-tracing algorithms.

Similar to the work of \cite{Aland2020}, this work provides a general non-iterative method suitable for any geometry.
It leverages the diffuse interface method's structured meshes to use far more computationally efficient \emph{distance transform} algorithms.
The algorithm is based on the combination of a distance transform and a \emph{kernel function}.
The distance transform approximates a scalar distance function (from the interface) $d(\bm{x})$.
It is then used to generate the continuous phase field by applying the user-specified kernel function $\phi(d)$ that defines the structure of the continuous interface in closed form.
The width of the continuous interface is controlled by a user-specified parameter $\lambda$.
Past work (\cite{Treeratanaphitak2021}), analyzed different options for the kernel function and found minimal differences in solution accuracy when the interface width was small with respect to domain size.
The results presented in this work use the basis/interpolant function itself as the kernel function, and the error function was also considered.
The full algorithm is summarized as follows:
\begin{enumerate}
    \item Given the geometry specified as a discretized surface (boundary faces, edges, and vertices), generate a structured quadrilateral mesh that fully encompasses the geometry.
    This is the nonconformal mesh.
    \item Define a temporary phase field $\eta(\bm{x})$ on the nonconformal mesh.
    \item Iterate through the geometry faces and set $ \eta $ to one at any mesh nodes within a reasonable tolerance of the face. This maps the border of the geometry over the nonconformal domain.
    \item Calculate the chessboard distance transform (\cite{Fabbri2008}) of every mesh point in the nonconformal domain from this border, $d(\bm{x}, \eta)$.
    \item Scale the distance transform by the user-specified diffuse interface scale $\lambda$.
    \item Apply the user-specified kernel function to the scaled distance transform to produce a smooth transition from zero to one to zero over the border region. $\lambda$ sets the width of this transition.
    \item Use a floodfill algorithm (\cite{Russ2015}) to set $ \eta $ to one in the interior of the border generated in step 3. This maps a binary representation of the geometry over the non-conformal domain.
    \item Combine the kernel function output and $ \eta $ to produce an phase field that varies from $-1$ outside of the border generated in step 3, to zero at the border, to one inside of the border.
    \item Rescale the phase field to run from zero to one. This is the final phase field $\phi(\bm{x})$.
\end{enumerate}

\begin{figure}[H]
   \begin{subfigure}[b]{0.32\linewidth}
     \includegraphics[width=\linewidth]{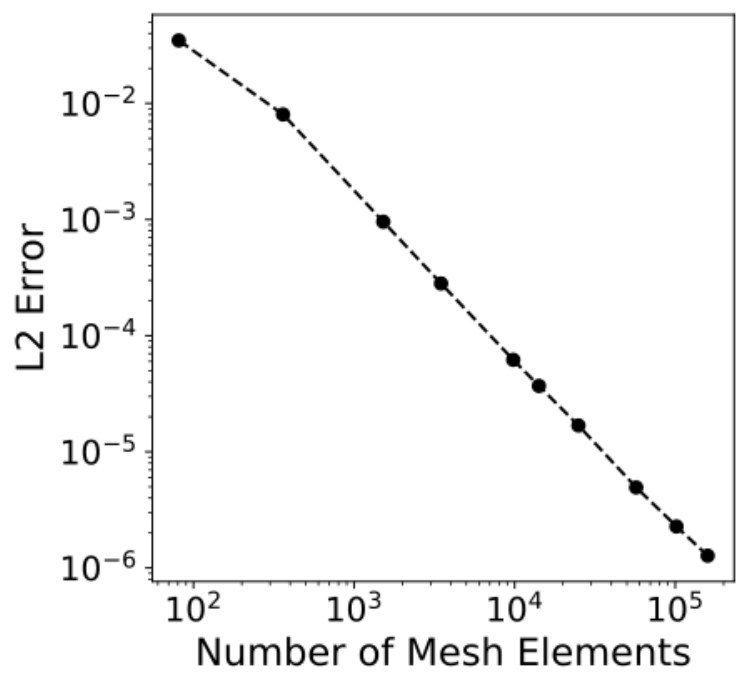}
      \caption{\label{fig:parameters:delta_x}}
   \end{subfigure}
	\hfill
 	\begin{subfigure}[b]{0.32\linewidth}
 		\includegraphics[width=\linewidth]{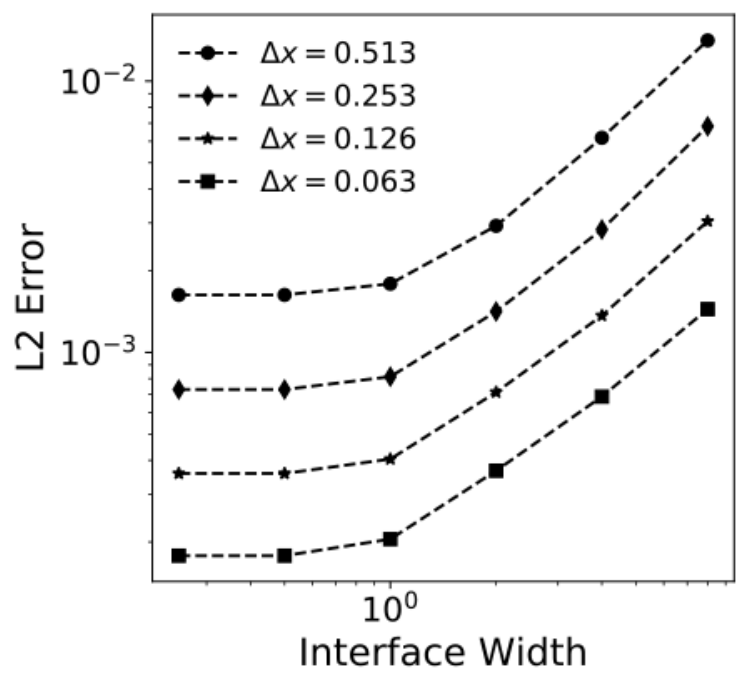}
     \caption{\label{fig:parameters:width}}
 	\end{subfigure}
   \hfill
 	\begin{subfigure}[b]{0.32\linewidth}
 		\includegraphics[width=\linewidth]{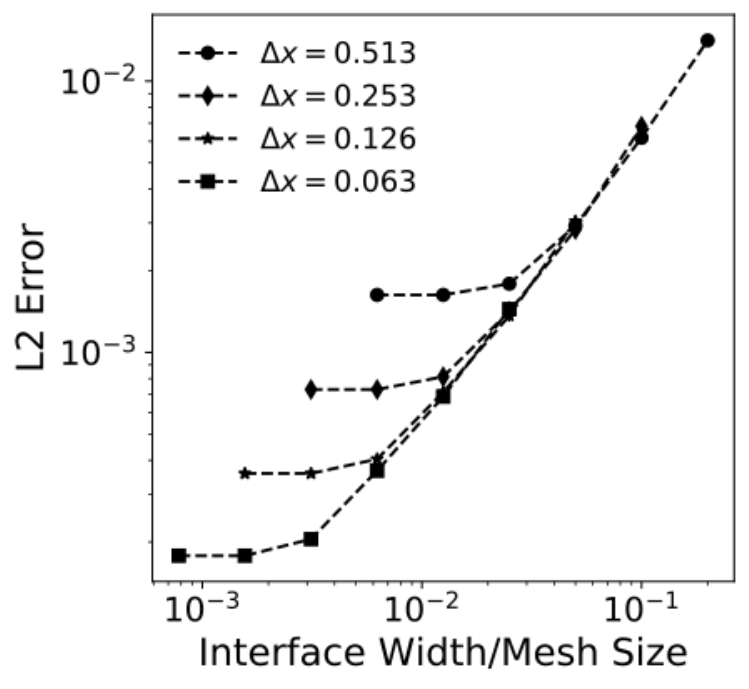}
     \caption{\label{fig:parameters:normalized_width}}
 	\end{subfigure}
 	\caption{The effects of (a) $\Delta x$, (b) the width of the diffuse interface, and (c) the width of the diffuse interface relative to the size of the domain on the accuracy of the diffuse interface method.}
 \end{figure}

There are two parameters inherent to the diffuse interface approximation to a domain boundary.
A third parameter is required when the Nitsche method is used to impose Dirichlet boundary conditions.
\begin{description}
	\item{Mesh Element Scale $\Delta x$:} This can be calculated for a given mesh dimension from the size of the nonconformal domain ($ L $) and the number of mesh elements ($ N $) along said dimension, $ \Delta x = L/(N-1)$. $\Delta x$ can be different for each different dimension of the nonconformal mesh.
    As shown in Figure \ref{fig:parameters:delta_x}, decreasing $ \Delta x $ increases the accuracy of the solution by increasing the order of the spatial approximation.
	\item{Diffuse Interface Scale $\lambda$:} This is a measure of the diffuseness of the interface.
    The width of the diffuse interface, defined as the number of mesh elements required for $ \phi $ to change from zero to one, can be calculated as $ w = 2 \lambda/\Delta x $.
    As shown in Figure \ref{fig:parameters:width}, $ \lambda $ affects the accuracy of the solution.
    Large values of $ \lambda $ extend the boundary condition constraints well into the interior of the complex geometry and generally decrease solution accuracy.
    The highest accuracy is typically observed when the diffuse interface is the same width as the mesh element scale, as this is the closest approximation to a sharp interface.
    For the same reasons, solution accuracy also improves as the width of the diffuse interface decreases relative to the scale of the nonconformal domain (see Figure \ref{fig:parameters:normalized_width}). However, when applying Neumann or Robin boundary conditions it may be preferable to use an interface width of several mesh elements as this smooths the gradient of $\phi$ at the boundaries and reduces numerical approximation errors.
	\item{Nitsche Stabilization Parameter $\beta$:} This parameter is required when using the Nitsche method to impose Dirichlet boundary conditions. As discussed in Section \ref{sec:background:Nitsche}, $ \beta $ can be obtained by solving an eigenvalue problem (\cite{Nguyen2018}) or calculated as $ \beta = 10 n^2 / \Delta x $, where $ n $ is the order of interpolant used (\cite{Evans2013, Benk2012, Ruess2013, Ruess2014}).
\end{description}

\begin{figure}[H]
   \begin{subfigure}[b]{0.32\linewidth}
     \includegraphics[width=\linewidth]{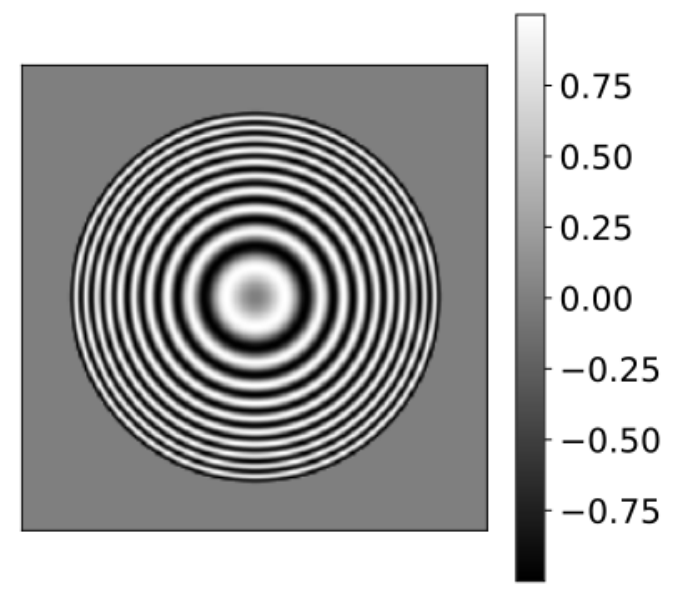}
      \caption{\label{fig:circle:exact}}
   \end{subfigure}
	\hfill
 	\begin{subfigure}[b]{0.32\linewidth}
 		\includegraphics[width=\linewidth]{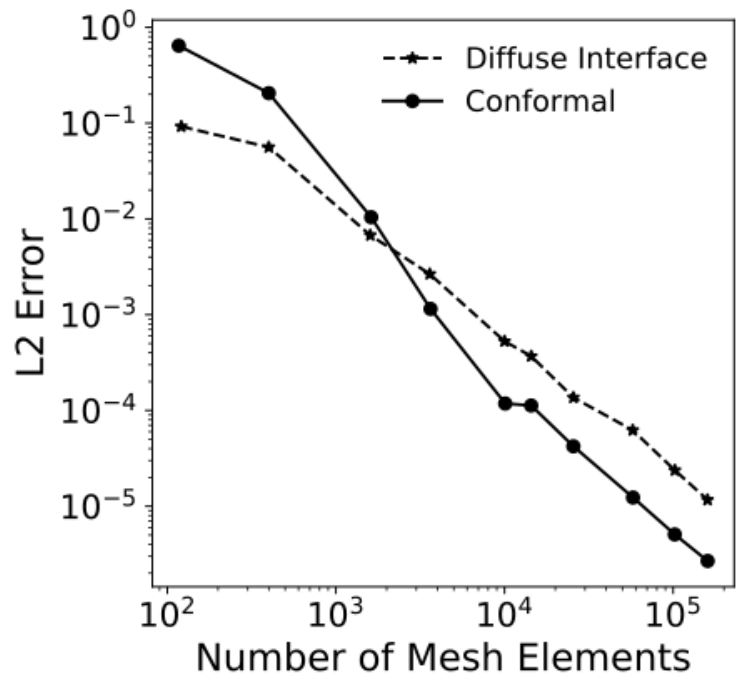}
     \caption{\label{fig:circle:dirichlet}}
 	\end{subfigure}
   \hfill
 	\begin{subfigure}[b]{0.32\linewidth}
 		\includegraphics[width=\linewidth]{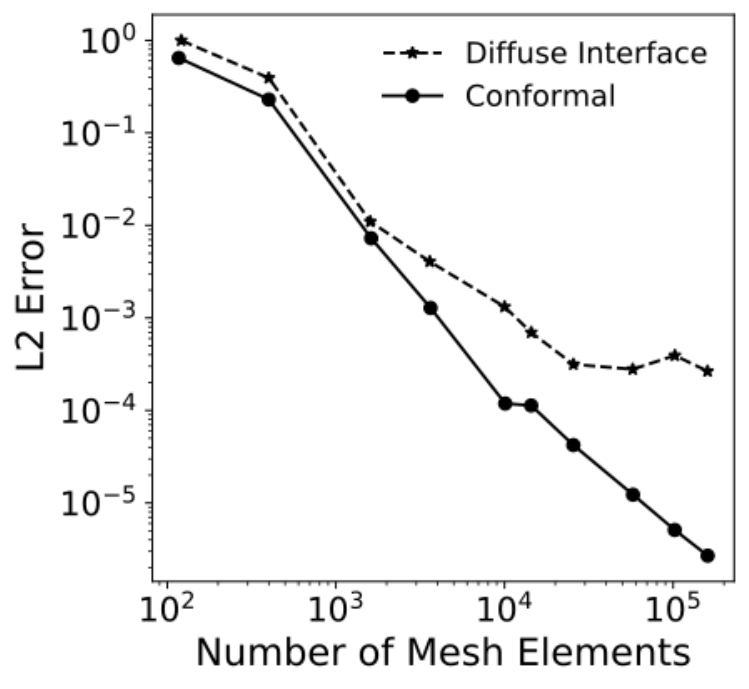}
     \caption{\label{fig:circle:neumann}}
 	\end{subfigure}
 	\caption{(a) surface plot of the exact solution (eqn. \ref{eqn:res:circle_exact}). Convergence behaviour of the conformal solution and diffuse interface method with (b) Dirichlet boundary conditions and (c) Neumann boundary conditions.}
 \end{figure}

The non-iterative method was applied to validation eqn. \ref{eqn:res:circle}.
The exact solution and its derivative (Figure \ref{fig:circle:exact}) were used to formulate uniform Dirichlet and Neumann boundary conditions respectively, and solutions were obtained from a standard conformal mesh and the diffuse interface method.
The convergence behaviour, global error versus number of mesh elements, of both methods is shown in Figures \ref{fig:circle:dirichlet}-\ref{fig:circle:neumann}.
In both cases, the conformal mesh reached a more accurate solution than the diffuse interface method.
In the case of uniform Neumann boundary conditions, the diffuse interface method also had a slightly lower convergence rate and converged at a smaller mesh size.
This is expected, since the diffuse interface method uses an inherently poorer interface approximation than a conformal mesh.

The diffuse interface method is expected to show significant performance benefits in the computation time needed to achieve a given level of accuracy.
However, for this simple two-dimensional circular geometry the computational overhead associated with the solver was sufficiently high to obscure any differences between the conformal mesh solution and the diffuse interface solution.
A more complex three dimensional problem is used in Section \ref{sec:results:validation} in order to evaluate this aspect of the performance of the method.

In the previous validation case, the geometry involved was a simple circle with a constant and bounded mean curvature, defined as,
\begin{equation}\label{fig:curvature}
    \kappa = -\frac{1}{2}\nabla_s\cdot\bm{n}
\end{equation}
where $\bm{n}$ is the unit normal to the boundary surface and $\nabla_s$ is the surface divergence.
The second validation case involves a square geometry, which is chosen due to the presence of points on the bounding surface in which the mean curvature is singular, that is, there are corners in the geometry.
This presents a more challenging geometry for the application of the diffuse interface method.

\begin{figure}[H]
   \begin{subfigure}[b]{0.32\linewidth}
     \includegraphics[width=\linewidth]{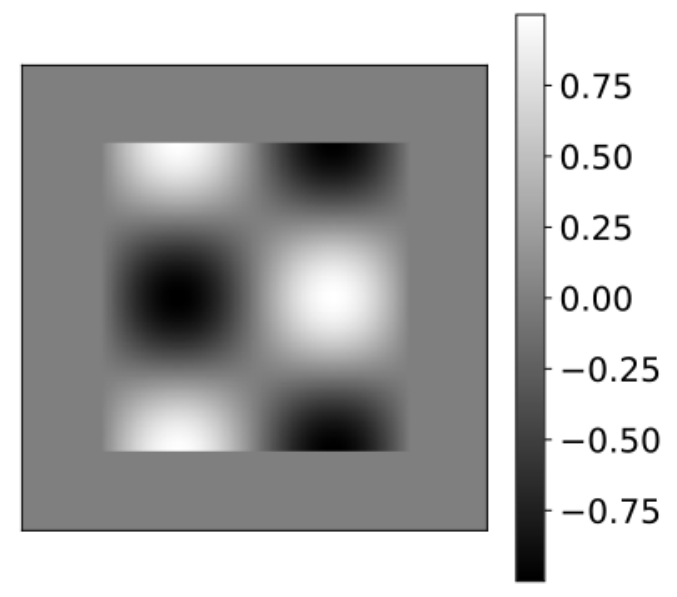}
      \caption{\label{fig:square:exact}}
   \end{subfigure}
	\hfill
 	\begin{subfigure}[b]{0.32\linewidth}
 		\includegraphics[width=\linewidth]{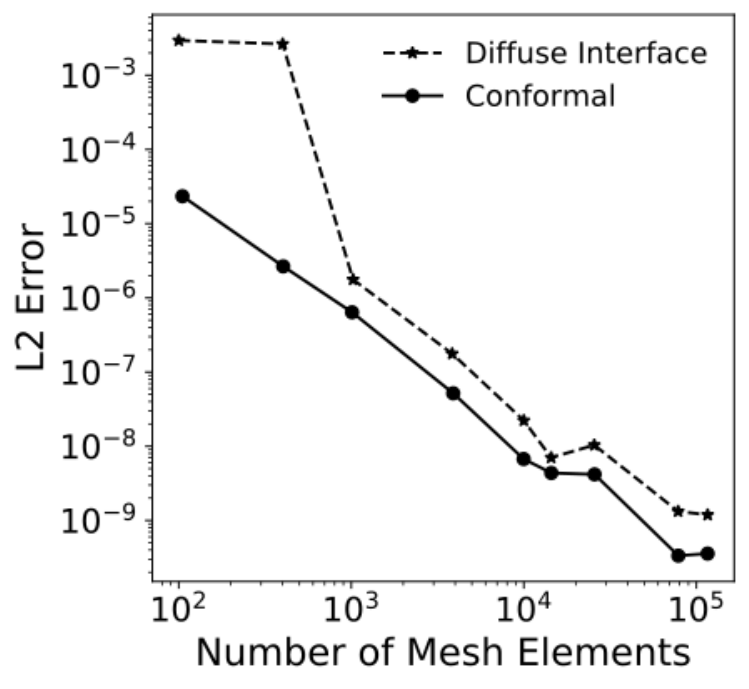}
     \caption{\label{fig:square:dirichlet}}
 	\end{subfigure}
   \hfill
 	\begin{subfigure}[b]{0.32\linewidth}
 		\includegraphics[width=\linewidth]{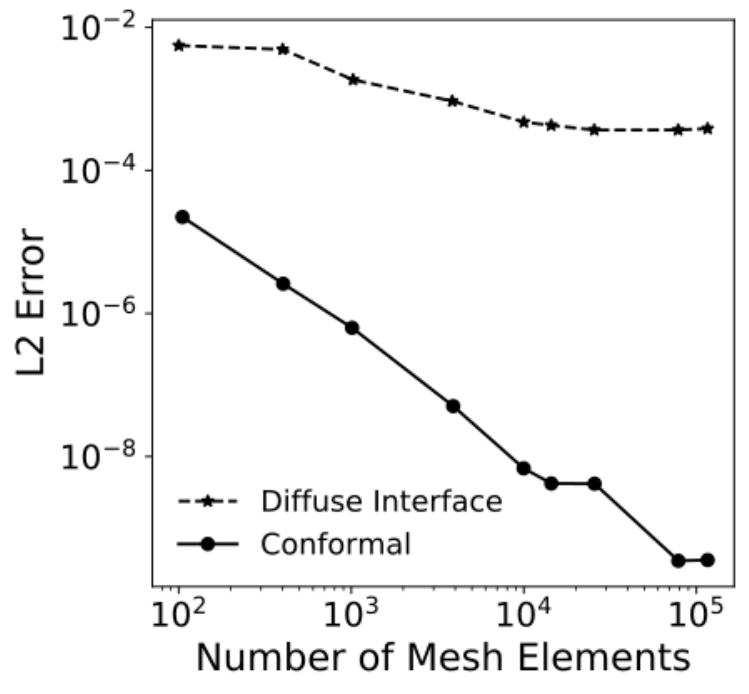}
     \caption{\label{fig:square:neumann}}
 	\end{subfigure}
 	\caption{(a) surface plot of the exact solution (eqn. \ref{eqn:res:square_exact}). Convergence behaviour of the conformal solution and diffuse interface method with (b) Dirichlet boundary conditions and (c) Neumann boundary conditions.}
 \end{figure}

The method was applied to validation eqn. \ref{eqn:res:square}, again for uniform Dirichlet and Neumann boundary conditions obtained from the exact solution and its derivative respectively (Figure \ref{fig:square:exact}).
The convergence behaviour, global error versus number of mesh elements, for the diffuse interface method and a standard conformal mesh are shown in Figures \ref{fig:square:dirichlet}-\ref{fig:square:neumann}.
Again, the conformal mesh solution shows better convergence behaviour in both cases.
Particularly for the uniform Neumann boundary conditions, it is significantly more accurate and has a significantly higher convergence rate than the diffuse interface method.
Moderate differences are expected, but clearly the presence of discontinuities in the curvature of the boundary has a further significant effect on the performance of the diffuse interface method.
The performance of the conformal mesh solution is also especially biased in this case, since the domain geometry (square) is conformal to the mesh geometry (rectangular).

\begin{figure}[h]
	\centering
	\begin{subfigure}[b]{0.49\textwidth}
		\includegraphics[width=\linewidth]{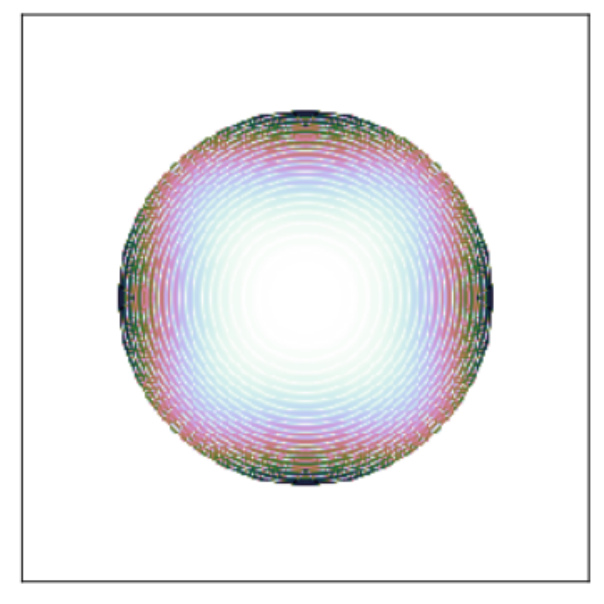}
		\caption{}
	\end{subfigure}%
	\hfill
	\begin{subfigure}[b]{0.49\textwidth}
		\includegraphics[width=\linewidth]{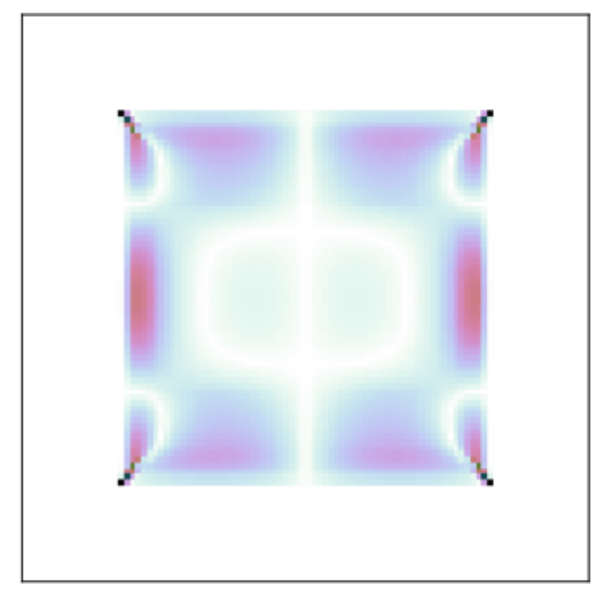}
		\caption{}
	\end{subfigure}%
	\vfill
	\begin{subfigure}[b]{0.49\textwidth}
		\includegraphics[width=\linewidth]{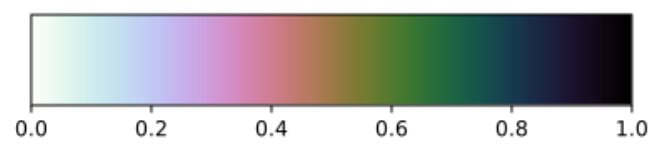}
	\end{subfigure}%
	\caption{Surface plots of the local error for the diffuse interface method applied to the (a) circlular geometry and (b) square geometry test cases. In both cases the error is normalized to lie on [0,1]. \label{fig:curvature:circle_vs_square}}
\end{figure}

The effects of geometry curvature on the diffuse interface method are further evident from the spatial distribution of error.
Figure \ref{fig:curvature:circle_vs_square} shows surface plots of the relative error for each of the two test cases.
Dark areas indicate regions of high error and the intensity is scaled by the maximum error in each figure domain.
In the case of the circle geometry, error is distributed relatively uniformly around the entire circular subdomain.
It does increase slightly in areas where the geometry curvature is more poorly approximated and in areas where the exact solution has significant spatial variation (outer region).
The square geometry, likewise, shows higher error along the two boundaries that intersect the greatest magnitude of the cosine wave.
However, the error is distributed nonuniformly with the majority of it located at the corners of the square, due to the singularities in the boundary curvature.

\subsection{Measures of diffuse interface quality based on mesh properties}\label{sec:results:metrics}

The previously presented results show that there is significant variation in the performance of the diffuse interface method depending on the attributes of the complex geometry.
Consequently, a set of simple \emph{predictive} scalar measures are developed to evaluate the accuracy of the method and guide the selection of user-defined scales (mesh and diffuse interface).
Based a simple scaling analysis, several dimensionless parameters are proposed, formulated to range from $0\rightarrow \infty$, where values close to zero predict that the diffuse interface method will perform well, while values approaching $\infty$ predict the opposite.
They serve as predictive measures of the potential accuracy of the diffuse interface method based on domain geometry and physical scales:
\begin{description}
    \item{Interface Scale Measure $\lambda_d / \lambda_i$:} This is the ratio of the diffuse interface scale, a user-specified parameter, to the physical scale of the interface.
    Apart from simulations under conditions approaching critical points \cite{Anderson1998}, this value will approach $\infty$ for most continuum mechanical models.
    \item{Mean Chord Scale Measure $\lambda_d / \lambda_g$:} This is the ratio of the diffuse interface scale to the mean scale of the geometry.
    \item{Minimum Chord Scale Measure $\lambda_d / \lambda_n$:} This is the ratio of the diffuse interface scale to the smallest of the necks present in the geometry.
    \item{Curvature Scale Measure $\lambda_d / \lambda_c$:} This is the ratio of the diffuse interface scale to the mean radius of curvature of the domain, excluding discontinuities in curvature (edges, points).
    \item{Discontinuity Scale Measure $\sum_i L_i / \lambda_d$:} This is the ratio of the length of the discontinuities in the curvature of the domain (for an edge the length of the edge, for a point $L = \lambda_d$) to the diffuse interface scale.
\end{description}

The various measures were calculated for the square and circle geometry test cases.
Theoretical expressions for mean chord length and radius of curvature are known for these simple geometries, but would not be known for a general complex geometry.
Thus, the chord lengths and radii of curvature are instead approximated using the boundary vertices of the CAD representations of each geometry.
As can be seen in Table \ref{tab:results:metrics}, the approximated measures agree well with the theoretical measures.
The minor discrepancies are due to the CAD not perfectly approximating the geometry boundaries.
The minimum chord scale measure is not given in either case, since neither domain contains necks.
Similarly, the circle geometry contains no discontinuities, so the discontinuity scale measure is zero.
In contrast, the square domain contains a discontinuity at each of the four corners.
The remaining portions of the square boundary are straight lines, resulting in a curvature scale measure of zero.
The measures are sensitive to the smoothness of the CAD representation.
For example, a sufficiently coarse CAD representation of a circle geometry could give a nonzero value for the discontinuity scale measure.
This is a desirable property, since the diffuse interface approximation of the geometry boundary is entirely dependent on the fidelity of the CAD representation.

\begin{table}[H]
\caption{Quality metrics for the two sample geometries.\label{tab:results:metrics}}
\begin{tabular}{|c|c|c|c|c|}
\hline
 & \multicolumn{2}{c|}{Circle} & \multicolumn{2}{c|}{Square} \\ \hline
Measure & Theory & Calculated & Theory & Calculated \\ \hline
Mean Chord Scale & $ 0.196 \lambda $ & $ 0.187 \lambda $ & $ 0.113 \lambda $ & $ 0.105 \lambda $ \\
Minimum Chord Scale & N/A & N/A & N/A & N/A \\
Curvature Scale & $ 0.25 \lambda $ & $ 0.251 \lambda $ & $ 0 $ & $ 0 $ \\
Discontinuity Scale & $ 0 $ & $ 0 $ & $ 4 $ & $ 4 $ \\ \hline
\end{tabular}
\end{table}

\subsection{Generalized diffuse interface method for spatially-varying boundary condition type}\label{sec:results:boundary}

The majority of experimentally and industrially relevant design problems involve spatially varying boundary conditions.
For example, standard models for pressure-driven flow through a conduit involve the imposition of boundary conditions on velocity which vary from no-slip conditions on the conduit walls (Dirichlet velocity and Neumann pressure condition) to nonzero pressure/velocity inlet and outlet conditions.
To date, the vast majority of applications of the diffuse interface method are restricted to uniform boundary conditions, or account for spatial variation in a non-generalized manner (\cite{Nguyen2017,Ramiere2007,Nguyen2018,Teigen2009,Aland2020,Benk2012,De2016,Mo2018}).
Consequently, a method is developed for the imposition of spatially-varying boundary conditions that is complementary to the method presented in the previous section for the generation of the diffuse interface itself.

Similar to the requirement that the bounding surface of the complex geometry must be specified, so too must the sub-surfaces for each individual boundary condition.
Given a set of $i$ sub-surfaces, the proposed method introduces a set of scalar fields $\{\phi_i\}$ in addition to the boundary phase field $\phi$. These are used to indicate the local boundary condition in the following general form,
\begin{equation}\label{eqn:results:nonuniform_bcs}
        \sum_i \phi \phi_i \left( A_i \bm{n}_i \cdot \bm{\nabla}_i T + B_i T + C_i\right) = 0
\end{equation}
The values of $\{A_i, B_i\}$ specify the type of boundary condition. $A_i, B_i \ne 0$ is a Robin boundary condition, $A_i = 0$ is a Dirichlet boundary condition, and $B_i = 0$ is a Neumann boundary condition.
Eqn. \ref{eqn:results:nonuniform_bcs} replaces the set of boundary conditions from the original formulation and is incorporated into the weak formulation as described in Section \ref{sec:background:DIM}.
In order to maintain stability for the imposition of Dirichlet boundary conditions the Nitsche reformulation is also used (\cite{Nitsche1971}), which reformulates the Dirichlet condition into a Robin condition as shown in Section \ref{sec:background:Nitsche}.

\begin{figure}[H]
	\centering
	\begin{subfigure}[b]{0.24\linewidth}
		\includegraphics[width=\linewidth]{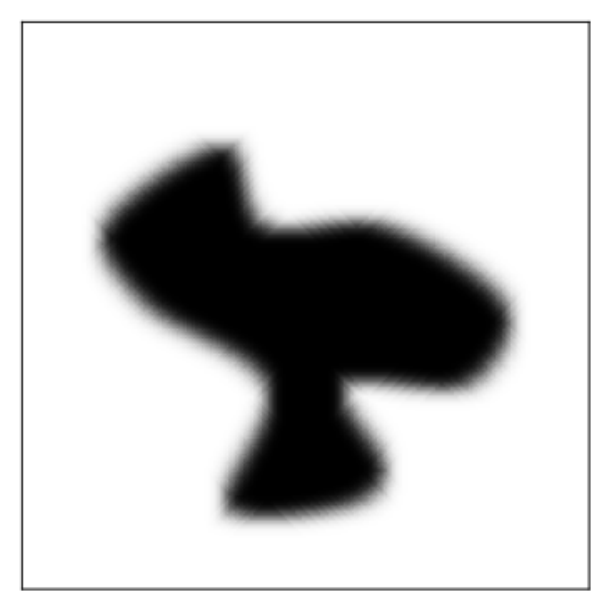}
		\caption{}
	\end{subfigure}%
	\hfill
	\begin{subfigure}[b]{0.24\linewidth}
		\includegraphics[width=\linewidth]{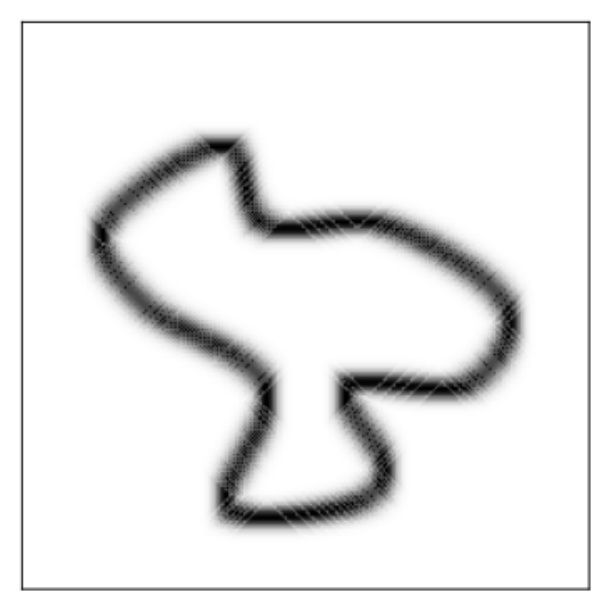}
		\caption{}
	\end{subfigure}%
	\hfill
	\begin{subfigure}[b]{0.24\linewidth}
		\includegraphics[width=\linewidth]{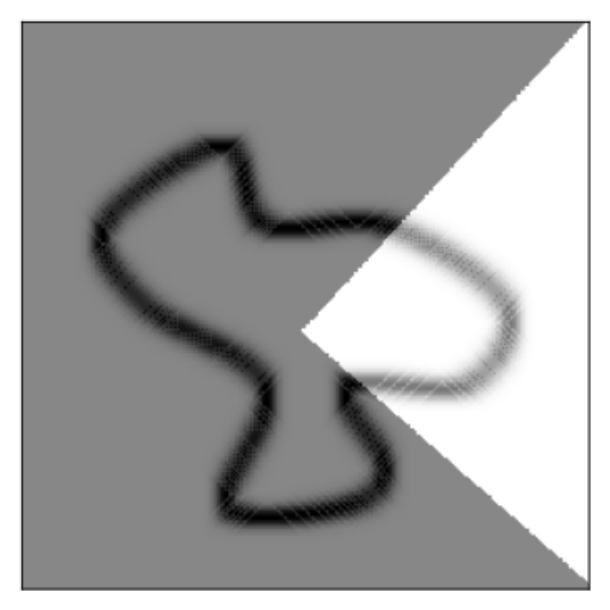}
		\caption{}
	\end{subfigure}%
	\hfill
	\begin{subfigure}[b]{0.24\linewidth}
		\includegraphics[width=\linewidth]{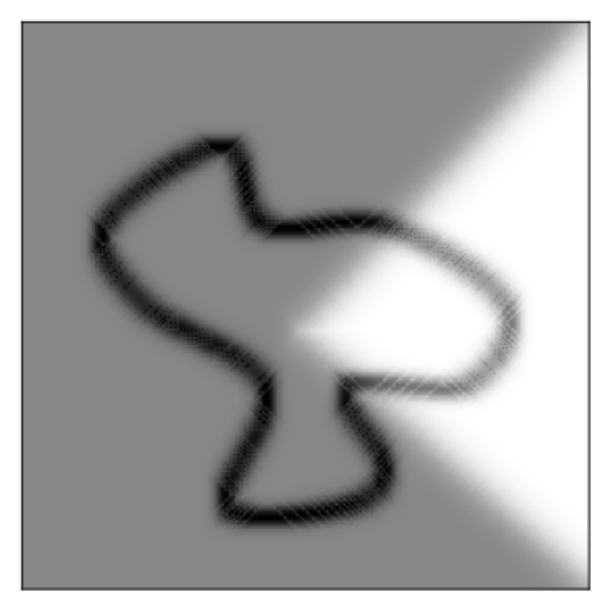}
		\caption{}
	\end{subfigure}%
		\caption{(a) $\phi$, (b) $\lvert \nabla \phi \rvert$ for the same complex geometry, (c) the mask used to apply a boundary condition to a section of $\lvert \nabla \phi \rvert$, and (d) the same mask if the different boundary conditions are made to diffuse into each other.\label{fig:results:nonuniform_bcs1}}
\end{figure}

The non-iterative method proposed is decomposed into (i) generation of the set of scalar fields $\{\phi_i\}$ and (ii) modification of the diffuse interface formulation to impose boundary conditions using this set.
The algorithm for generation of the scalar fields is summarized as follows and demonstrated for a two-dimensional domain in Figure \ref{fig:results:nonuniform_bcs1}:
\begin{enumerate}
    \item Given the geometry specified as a discretized surface (boundary faces, edges, and vertices), identify sub-sets of the discretized surface associated with each of the sub-surfaces on which an individual boundary condition is to be imposed.
    \item Calculate the centroid of the discretized geometry or use a user-specified location.
    \item Define a set of scalar fields $\{\phi_i\}$ on the nonconformal mesh. In two dimensions, each field is one at every mesh node inside the solid angle extending from the centroid through both bounding vertices to the edges of the nonconformal mesh. In three dimensions this is the cone extending from the centroid through all boundary section boundary vertices to the edges of the nonconformal mesh. The fields are zero everywhere else, resulting in masks over only specific regions of the domain.
    \item Optionally (if specified by the user), apply a distance transform weighted by $\lambda_{overlap}$ to the edge of each mask, then apply the kernel function to the result. This results in a diffuse transition between neighbouring boundary conditions.
\end{enumerate}

\begin{figure}[H]
	\centering
	\begin{subfigure}[b]{0.49\textwidth}
		\includegraphics[width=\linewidth]{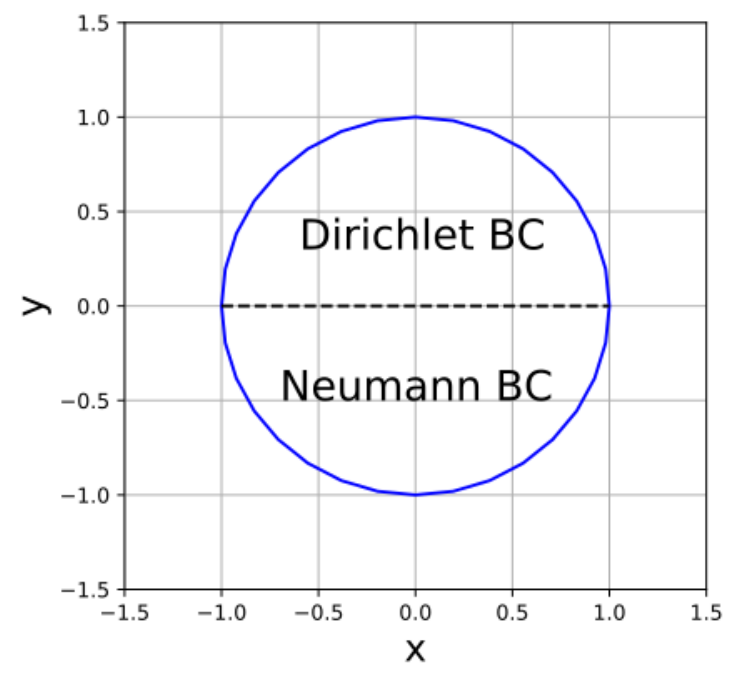}
     \caption{}
	\end{subfigure}
	\hfill
	\begin{subfigure}[b]{0.49\textwidth}
		\includegraphics[width=\linewidth]{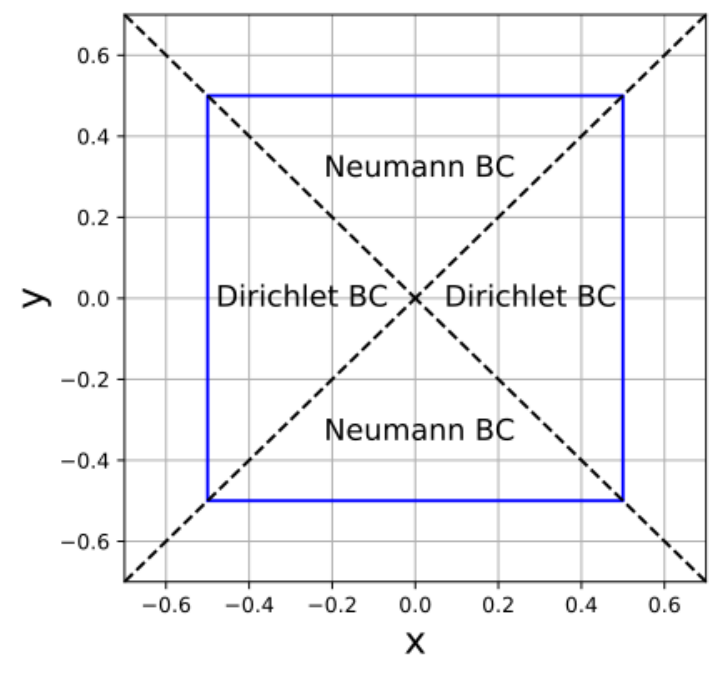}
     \caption{}
	\end{subfigure}\\
 	\begin{subfigure}[b]{0.49\linewidth}
 		\includegraphics[width=\linewidth]{circle_ngsolve_nd}
     \caption{}
 	\end{subfigure}
 	\hfill
 	\begin{subfigure}[b]{0.49\linewidth}
 		\includegraphics[width=\linewidth]{square_ngsolve_nd}
     \caption{}
 	\end{subfigure}\\
 	\begin{subfigure}[b]{0.49\linewidth}
 		\includegraphics[width=\linewidth]{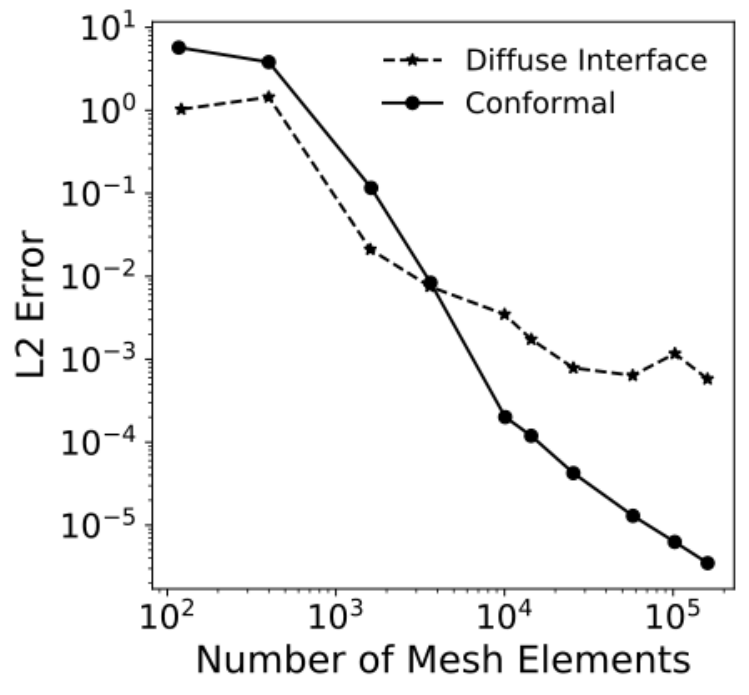}
     \caption{}
 	\end{subfigure}
   \hfill
 	\begin{subfigure}[b]{0.49\linewidth}
 		\includegraphics[width=\linewidth]{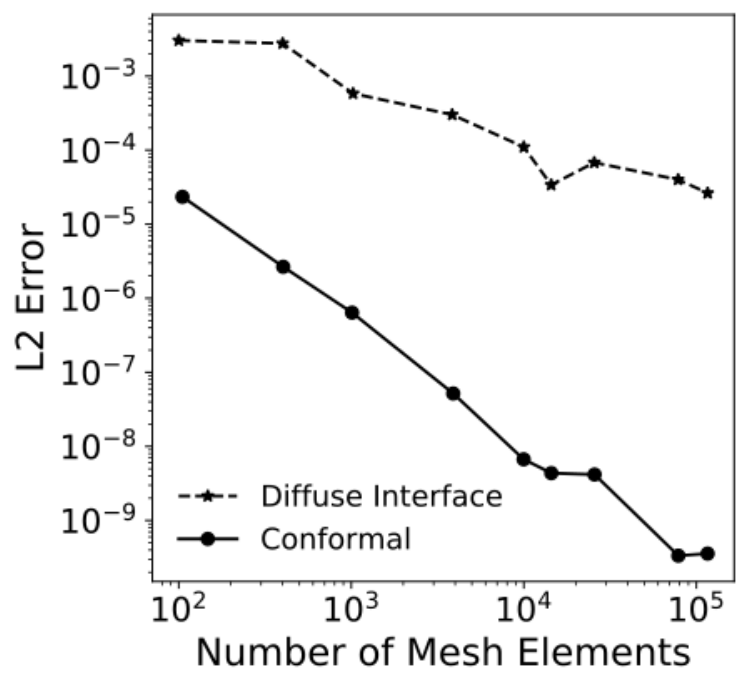}
     \caption{}
 	\end{subfigure}\\
 	\caption{The placement of the different boundary conditions on (a) the circle geometry and (b) the square geometry. (c,d) exact solutions (eqns. \ref{eqn:res:circle_exact} and \ref{eqn:res:square_exact}) and (e,f) convergence behaviour for the diffuse interface method with mixed Dirichlet and Neumann boundary conditions.\label{fig:results:nonuniform_bcs2}}
 \end{figure}

The method was applied to the same two-dimensional test cases used in section \ref{sec:results:generation}, with eqns. \ref{eqn:res:circle}-\ref{eqn:res:square} formulated for mixed Dirichlet and Neumann boundary conditions.
The placement of the different boundary conditions is shown in Figure \ref{fig:results:nonuniform_bcs2}a,b. Figure \ref{fig:results:nonuniform_bcs2}e,f then shows the convergence behaviour of the diffuse interface method with spatially-varying boundary conditions, compared to solutions on standard conformal meshes.
As expected, the conformal solutions show better convergence behaviour, higher accuracy, except on very coarse meshes, and higher rates of convergence.
Indeed, in both cases, the diffuse interface method shows similar trends and accuracy for spatially-varying boundary conditions as for uniform conditions.
This shows, qualitatively, that the additions to the diffuse interface method for imposing spatially-varying boundary conditions have no significant effect on the performance of the method.

\subsection{Application to a heat transfer problem}\label{sec:results:validation}

The methods presented in Sections \ref{sec:results:generation}-\ref{sec:results:boundary} were then used for a realistic application; evaluation of the heat transfer performance of a three dimensional LED heat sink design.
This demonstrates the potential performance of the diffuse interface method with mixed boundary conditions on a complex geometry and allows for a full comparison with standard conformal mesh solutions, including a timing comparison.

A CAD representation of the design is shown in Figure \ref{fig:results:heat_sink}.
The heat sink itself has a diameter of $\SI{12}{cm}$ and a height of $\SI{5.59}{cm}$.
It is assumed to be pure aluminum, with a heat capacity of $\SI{910}{J kg^{-1} K^{-1}}$ and a thermal conductivity of $\SI{205}{W m^{-1} K^{-1}}$ (\cite{Young2008}).
The bottom of the heat sink is in contact with three rectangular LEDs, each of which output $\SI{100}{W}$ and which have areas of $\SI{6.45}{cm^2}$, $\SI{1.61}{cm^2}$, and $\SI{9.68}{cm^2}$.
The remaining finned surfaces experience natural convection in air, with a convective heat transfer coefficient of $\SI{10}{W m^{-2} K^{-1}}$ and an environmental temperature of $\SI{298}{K}$.
Steady-state heat transfer within the heat sink is modelled using the following approximations:
\begin{enumerate}
    \item Temperature-independent material properties.
    \item Convective heat transfer boundary conditions for the heat sink/air interface (Figure \ref{fig:results:heat_sink}a),
    \begin{equation}
        \bm{n}\cdot\bm{q} = h(T - T_{\infty})
    \end{equation}
    which corresponds to a Robin boundary condition.
    \item Heat flux boundary conditions for the LED/heat sink interface (Figure \ref{fig:results:heat_sink} b), which corresponds to a Neumann boundary condition.
\end{enumerate}
The complexity of the CAD geometry can be assessed with the mesh quality metrics described in Section \ref{sec:results:metrics}.
The interface scale measure is $\infty$ since the physical interface is assumed to be infinitesimally thin.
The mean chord scale measure is $\lambda / \SI{0.358}{cm}$ and the minimum chord scale measure is $\lambda / \SI{2.941}{cm}$.
The curvature scale measure is $0$ as the heat sink has many flat edges.
Finally, the discontinuity scale measure is $\SI{253.7}{cm}/ \lambda$ and is only comprised of edge discontinuities.
Comparing these values to Table \ref{tab:results:metrics} confirms that the LED heat sink is far more difficult for the diffuse interface method to approximate than a circle or square.
All of the diffuse interface solutions use a diffuse interface thickness of $\SI{0.165}{cm}$, so the mesh quality metrics are then $\infty$, $\SI{0.461}{cm^{-1}}$, $\SI{0.0561}{cm^{-1}}$, $0$, and $\SI{1537.6}{cm}$ respectively.

All conformal meshes are completely unstructured and were generated with Netgen/NGSolve (\cite{ngsolve}), using its automated mesh optimization tools.
This meshing and the simulations were performed on one core of a standard laptop under identical power usage conditions for both the conformal solutions and the diffuse interface solutions.
The range of mesh sizes was chosen, based on total simulation time, to extend from the coarsest (and fastest to generate) possible meshes to the beginnings of a mesh-independent diffuse interface solution.
This gave a wider range of mesh sizes for the diffuse interface method compared to the conformal solutions, as Netgen/NGSolve required a much higher minimum number of mesh elements to generate stable conformal meshes. This demonstrates one benefit of the diffuse interface method for geometries with a wide range of feature sizes.
The conformal meshes are constrained by the minimum feature size of the geometry, either resulting in highly skewed elements within coarse meshes, or requiring finer meshes for stability and reasonable element quality. In contrast, the diffuse interface method is not constrained by the geometry when meshing even if the phase field poorly approximates the geometry boundary.

\begin{figure}[H]
	\centering
	\begin{subfigure}[b]{0.49\linewidth}
		\includegraphics[width=\linewidth]{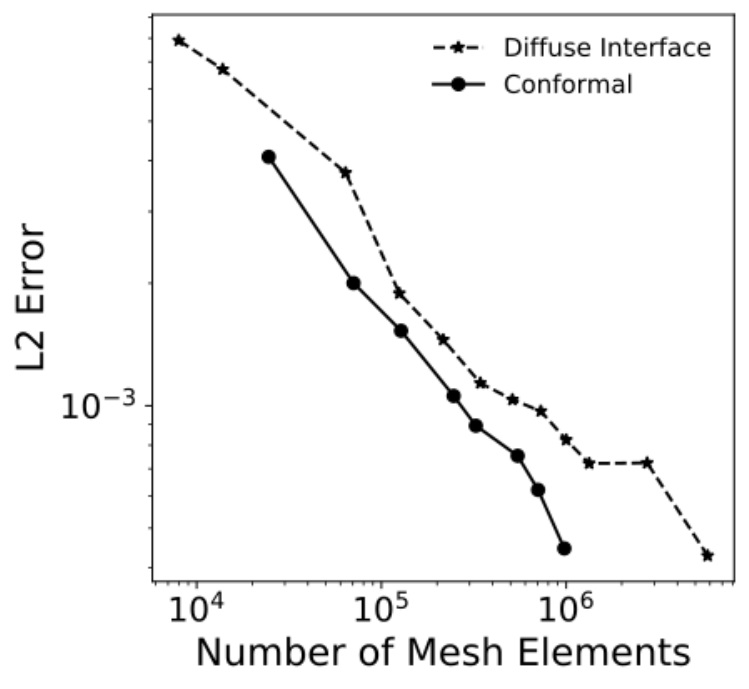}
		\caption{}
	\end{subfigure}
	\hfill
	\begin{subfigure}[b]{0.49\linewidth}
		\includegraphics[width=\linewidth]{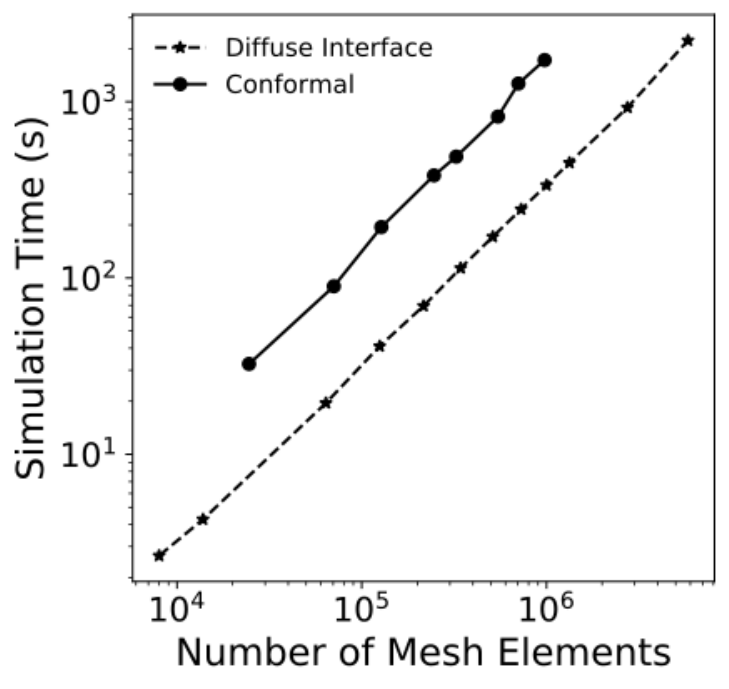}
		\caption{}
	\end{subfigure}
	\\
	\begin{subfigure}[b]{0.49\linewidth}
		\includegraphics[width=\linewidth]{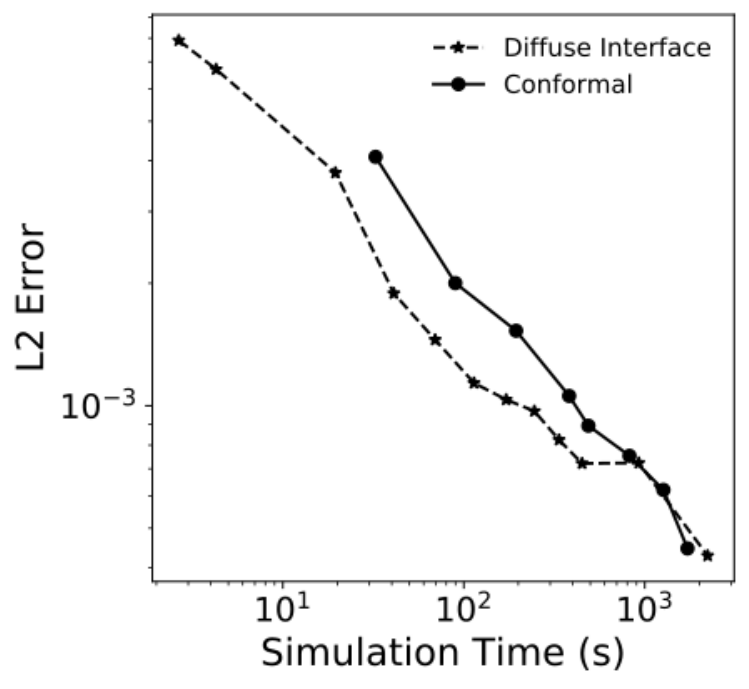}
		\caption{}
	\end{subfigure}
	\hfill
	\begin{subfigure}[b]{0.49\linewidth}
		\includegraphics[width=\linewidth]{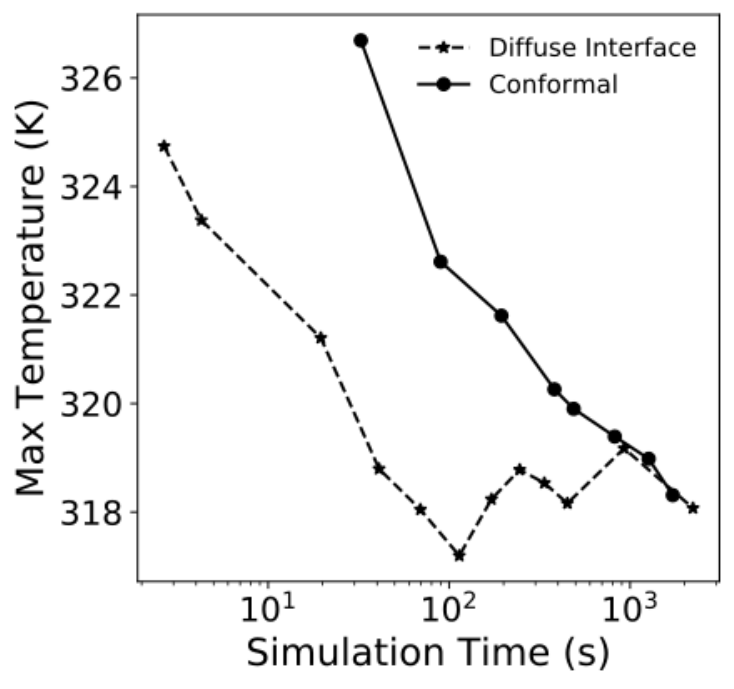}
		\caption{}
	\end{subfigure}
	\caption{(a) convergence behaviour of the conformal solution and the diffuse interface method, (b) timing comparison of the conformal solution and the diffuse interface method, (c) timing compared to error, and (d) timing compared to the maximum temperature of the heat sink.\label{fig:results:heat_sink2}}
\end{figure}

Figure \ref{fig:results:heat_sink2}a-d compares the performance of the two methods on several criteria: global error, number of mesh elements, total simulation time, and maximum temperature of the heat sink.
As no analytical solution exists, both the conformal mesh and the diffuse interface numerical solutions were compared to a converged (mesh-independent) conformal solution with an error tolerance of $\SI{1e-4}{}$.
As shown in Figure \ref{fig:results:heat_sink2}a, the standard conformal mesh approach gives higher accuracy than the diffuse interface method for a given number of mesh elements.
This is expected, as the diffuse interface method approximation of a geometry boundary is, in general, less accurate than that of a conformal mesh.
However, as shown in Figure \ref{fig:results:heat_sink2}b, the total simulation time required by the diffuse interface method scales significantly better than that of the standard approach.
This is predominantly due to the increased computation time required to generate an unstructured conformal mesh compared to a structured grid.
The use of solvers optimized for structured meshes would further increase this timing difference, but was not feasible for this work.
Accounting for total simulation time (see Figure \ref{fig:results:heat_sink2}c), the diffuse interface method provides similar accuracy for a lower computation time, until very fine meshes are required.
As shown in Figure \ref{fig:results:heat_sink2}d, the diffuse interface method also gives reasonable estimates of key design parameters (in this case the maximum temperature of the heat sink) significantly faster than the standard conformal solutions.
This is desirable for engineering design activities where screening of a large number of candidate designs is only feasible for a constrained amount of simulation time.

\section{Conclusions}\label{sec:conclusions}

A comprehensive method for simulation of heat transfer processes using the diffuse interface to capture complex domain boundaries was presented and compared to a traditional conformal mesh approach.
This method was developed as an alternative to traditional conformal mesh simulations for screening of a large number of design variations as part of a design process.
It reduces the computational complexity and avoids the user-interaction required by conformal mesh creation.
However, the reduced computational complexity enabled by the diffuse interface method comes at the cost of accuracy.

The methods presented and demonstrated in this work include automated non-iterative generation of phase fields from CAD geometries (Section \ref{sec:results:generation}) and an extension of the diffuse interface method to impose mixed boundary conditions (Section \ref{sec:results:boundary}).
Additionally, simple measures of diffuse interface quality based on CAD and mesh properties were presented and demonstrated to provide an understanding of the performance of the diffuse interface method.

The diffuse interface method was then applied to a realistic heat transfer problem, evaluation of the performance of an LED heat sink, and compared to the traditional conformal mesh approach..
It is found to enable reasonable accuracy at an order-of-magnitude reduction in simulation time or comparable accuracy for equivalent simulation times.
This supports its use for screening large numbers of design variations as part of the design heat transfer processes and its future extension to fluid dynamics, mass transfer, and multiphysics processes.

\section*{Acknowledgements}

This research was supported by the Natural Sciences and Engineering Research Council (NSERC) of Canada, the Ontario Ministry of Ministry of Research, Innovation and Science, and Compute Canada.

\typeout{}
\bibliographystyle{model4-names}\biboptions{authoryear}
\bibliography{computational,multiphase,comphys}
\end{document}